\documentclass{aa}  
\bibpunct{(}{)}{;}{a}{}{,} % to follow the A&A style
\usepackage[varg]{txfonts}
\pdfoutput=1 %for arXiv submission
\usepackage{amsmath,amssymb,amstext}
\usepackage[T1]{fontenc}
\usepackage{color}

\begin{document}

\title{The Fermi bubbles from stochastic acceleration of electrons in a Galactic outflow}

\author{P.~Mertsch\inst{\ref{inst1},\ref{inst2},\ref{inst3}}
\and
V.~Petrosian\inst{\ref{inst3},\ref{inst4}}
}

\institute{
Institute for Theoretical Physics and Cosmology (TTK), RWTH Aachen University, Sommerfeldstr. 16, 52074 Aachen, Germany\\
\email{pmertsch@physik.rwth-aachen.de}\label{inst1}
\and
Niels Bohr International Academy, Niels Bohr Institute, Blegdamsvej 17, 2100 Copenhagen, Denmark\label{inst2}
\and
Kavli Institute for Particle Astrophysics \& Cosmology, 2575 Sand Hill Road, M/S 29, Menlo Park, CA 94025, USA\label{inst3}
\and
Department of Physics and Applied Physics, Stanford University, Stanford, CA 94305, USA\label{inst4}
}

\abstract{
The discovery of the Fermi bubbles---a huge bilobular structure seen in GeV gamma-rays above and below the Galactic center---implies the presence of a large reservoir of high energy particles at $\sim 10 \, \text{kpc}$ from the disk. The absence of evidence for a strong shock coinciding with the edge of the bubbles, and constraints from multi-wavelength observations point towards stochastic acceleration by turbulence as a likely mechanism of acceleration. We have investigated the time-dependent acceleration of electrons in a large-scale outflow from the Galactic centre. For the first time, we present a detailed numerical solution of the particle kinetic equation that includes the acceleration, transport and relevant energy loss processes. We also take into account the addition of shock acceleration of electrons at the bubble's blast wave. Fitting to the observed spectrum and surface brightness distribution of the bubbles allows determining the transport coefficients, thereby shedding light on the origin of the Fermi bubbles.
}


\maketitle

%------------------------------------------------------------------------------------------------------------------------------------------
%------------------------------------------------------------------------------------------------------------------------------------------
%------------------------------------------------------------------------------------------------------------------------------------------
\section{Introduction}

The detection of the \textit{Fermi} bubbles--a huge bi-lobular
structure seen in GeV gamma-rays--is certainly one of the great
discoveries made with the \textit{Fermi}-LAT instrument. Due to their
position on the sky (see below), they are likely emanating from the
Galactic centre and the most speculated about sources are the
supermassive black hole at the Galactic centre and star formation/star
burst in the Galactic centre region. These processes shape Galactic
structure on the largest scales and as such the Fermi bubbles allow us
to study Galactic feedback in our own backyard. Furthermore, given
their prominence in gamma-rays, they are an important arena for
studies of sources of diffuse GeV emissions, like searches for signals
from self-annihilation or decay of dark matter. Finally, the
production of the gamma-rays and the acceleration of the underlying
particles are of astrophysical interest in itself.

%------------------------------------------------------------------------------------------------------------------------------------------
%------------------------------------------------------------------------------------------------------------------------------------------
\subsection{Observational properties}

Originally the Fermi bubbles were observed in a search
\citep{Dobler:2009xz} for the gamma-ray counterpart of a microwave
excess seen from the inner Galaxy
\citep{Finkbeiner:2003im,Dobler:2007wv,Ade:2012nxf}. A more detailed
analysis \citep{Su:2010qj} unveiled some surprising properties that
were later largely confirmed by \citet{Fermi-LAT:2014sfa}. In the
following we summarise the most important observational properties of
the Fermi bubbles in gamma-rays.

\paragraph{Geometry:} The Fermi bubbles are approximately centered at zero
Galactic longitude, symmetric about the Galactic plane, $50^\circ$
wide in longitude with each bubble extending up to $50^\circ$ in
latitude, see, e.g.\ Fig.~22 of \citet{Fermi-LAT:2014sfa}. On these
scales, they constitute the first evidence for an outflow from the
Milky Way. (On smaller scales, there had previously been evidence in
X-rays in an X-shaped feature around the Galactic centre.) The
bubbles' symmetry about the Galactic plane and their being centred
around zero longitude imply an origin at Galactic centre (distance
$d_{GC} \simeq 8.5 \, \text{kpc}$). A wind with a constant speed of
$1000 \, \text{km} \, \text{s}^{-1}$ would need about $9.9 \,
\text{Myr}$ to expand into a bubble of size $\sim d_{GC} \tan
50^{\circ} \simeq 10.1 \, \text{kpc}$, modulus projection effects: At
a latitude of $50^{\circ}$, we might be seeing the limb-brightened
edge of a bubble of radius $d_{GC} \sin 50^{\circ} \simeq 6.5 \,
\text{kpc}$, thus reducing the time-scale to $6.4 \, \text{Myr}$. Note
that because the eastern edge of the northern bubble is very close to
the position of the North-polar spur, which is part of the the radio
Loop~I. Initially, this led to claims of the bubbles being associated
with the Loop~I structure \citep{Casandjian:2009wq}.

\paragraph{Spectrum:} The gamma-ray flux shows a hard spectrum, mostly $\propto
E^{-2}$ and extending from a few hundred MeV up to a few hundred GeV,
see, e.g.\ Fig.~18 of \citet{Fermi-LAT:2014sfa}. At lower energies,
the spectrum is significantly harder, and at high energies there is
evidence for a spectral softening or an exponential cut-off. This
spectral shape immediately invites speculation about its physical
origin, i.e. whether the gamma-rays are of leptonic (from
inverse-Compton scattering) or hadronic ($\pi^0$ decay) origin. (Given
the estimates of the physical conditions, inside the bubbles, see
below, bremsstrahlung is most likely negligible.)

While the spectral shoulder around a few hundred MeV determined in the
earlier analysis \citep{Su:2010qj} seemed to be well fit by the
kinematic feature from $\pi^0$ decay, the new best-fit spectrum
appears to be extending to lower energies. Likely, a hadronic model
needs to have a spectral break (a steeper spectrum of the underlying
protons) at lower energies. This is in addition to the required
spectral break or cut-off at high energies. The physical origin of
these breaks is \textit{a priori} unclear.

In leptonic models these breaks are easily explained. The
inverse-Compton spectrum is naturally rather hard: In the Thomson
regime, a gamma-ray spectrum $\propto \epsilon^{-s}$ with ($s\sim 2)$
with a cutoff at $\epsilon_{\rm cut}$ of a few hundred GeV can be
produced by an electron spectrum $\propto E^{-\Gamma}
\mathrm{e}^{-E/E_{\mathrm{cut}}}$ with $\Gamma = 2s-1\sim 3$ and
$E_{\mathrm{cut}}=/m_{\mathrm{e}} c^2\sqrt{\epsilon/\epsilon_{\rm
    soft}}\sim 1,500$ GeV for soft photon energies of $\epsilon\sim 1$
eV.  Note that these estimates are strictly only valid in the Thomson
regime. In the numerical computations, however, we have used the full
Klein-Nishina cross-section and taken into account the relativistic
corrections.

\paragraph{Surface brightness.} The surface brightness shows little 
variation over the bubbles, but has sharp edges as can already be seen
in the residual map, cf.~e.g.~Fig.~29 of
\citet{Fermi-LAT:2014sfa}. More quantitatively, this is evidenced by
profiles of the gamma-ray flux across the bubble edge, shown e.g. in
Fig. 22 of \citet{Fermi-LAT:2014sfa}. There is clearly a jump in
intensity from a value close to zero (after template subtraction)
outside to a relatively constant value inside the bubbles. In fact,
the only substructure seen is a rather large enhancement of emissivity
in the east of the southern bubble, called the ``cocoon'', the origin
of which is of yet unknown. There have also been claims of evidence
for a narrow and extendend, jet-like feature \citep{Su:2012gu},
however, the analysis by the Fermi collaboration
\citep{Fermi-LAT:2014sfa} has found this feature not to be
significant.

The flat surface brightness and sharp edges are one of the most
puzzling features of the bubbles. The sharp edges require an efficient
confinement of the gamma-ray producing particles and the flat surface
brightness requires a peculiar distribution of volume
emissivity. Idealising each bubble as a spherically symmetric volume
with outer radius $R$, only an emissivity that varies with radius $r$
as $1 / \sqrt{1 - (r/R)^2}$ will give a flat surface brightness and
sharp edges.

\paragraph{Spectral uniformity.} The bubbles show similar morphologies in
different energy bins ranging from $1$ to $500 \, \text{GeV}$ (see
e.g.~Fig.~22 of \citet{Fermi-LAT:2014sfa}) or equivalently the
spectrum is uniform in different parts of the bubbles. Specifically,
the gamma-ray spectrum has been analyzed in different latitude bands
and the spectrum in the bubble edge region and the interior have been
compared: For the latitude bands, no variation has been found above
and below $\pm 10^{\circ}$. Between $-10^{\circ}$ and $+10^{\circ}$
there is an excess at the Galactic Centre \citep{Hooper:2013rwa},
likely with a spherical symmetry, and its connection to the Fermi
bubbles is unclear at this point
\citep{TheFermi-LAT:2017vmf}. Furthermore, no variation between the
edge region and the interior was found \citep{Su:2010qj} (but see also
\citet{Keshet:2016fbq}).

The spectral uniformity is also very surprising for such an extended
structure.  Leptonic models in particular would be expected to lead to
some variation, depending on the region of energizing of the
high-energy electrons. This is due to cooling losses by synchrotron
radiation and inverse-Compton emission. A conservative estimate of the
cooling time is $\tau_{\text{cool}} = 6.74 \times 10^8 \text{yr} \,
(E/\text{GeV})^{-1} ((u_B + u_{\text{CMB}})/ (0.486 \, \text{eV}
\text{cm}^{-3}) )^{-1}$, for magnetic fields and radiation fields of
energy densities $u_B = 0.224 \, \text{eV} \, \text{cm}^{-3} (B / 3
\mu\text{G})^2$ and $u_{\text{CMB}} = 0.262 \, \text{eV} \,
\text{cm}^{-3}$, respectively, i.e. of the same order as the bubble
age for $100 \, \text{GeV}$ electrons. Therefore, electrons energised
in the Galactic plane will be subject to considerable cooling while
travelling out into the bubble volume. This results in softer spectra
at larger distances from the Galactic centre and thus a softer
gamma-ray spectrum at higher latitudes. In addition, the energy
densities in the radiation backgrounds that the electron
inverse-Compton scatter on should be varying with distance from the
disk: While the CMB is of course spatially uniform, the energy
densities in both the optical/UV and the infrared backgrounds should
become smaller further away from the disk. The fact that this is not
observed imply that the variation in the radiation backgrounds must be
counter-balanced by a variation in the electron spectrum to some
degree.

%------------------------------------------------------------------------------------------------------------------------------------------
\subsection{Hints}

While the discovery of the Fermi bubbles was certainly a surprise, it
was not the first hint at the presence of Galaxy-scale
outflows. Kiloparsec-scale outflows have been observed for starburst
galaxies, e.g. in ionised gas. Even in our own Milky Way, there had
been hints at the presence of a Galaxy-scale outflow, possibly
connected with high-energy cosmic rays: Observations in soft X-rays,
most notably from ROSAT, showed signs of an x-shaped feature,
interpreted as evidence of a biconical outflow in analogy with
structures seen in other galaxies.

The presence of a population of high-energy cosmic ray electrons was
already hinted at by the microwave haze, an excess of microwaves from
the Galactic centre, pointing at a similarly hard electron spectrum
\citep{Finkbeiner:2003im,Dobler:2007wv,Ade:2012nxf}. (Note, however,
the possible influence of systematic effects due to template
subtraction \citep{Mertsch:2010ga}.) The search for a counterpart of
the microwave haze in gamma-rays was in fact what motivated the first
study that lead to the discovery of the Fermi
bubbles~\citep{Dobler:2009xz}.

%------------------------------------------------------------------------------------------------------------------------------------------
\subsection{Other constraints}
\label{sec:other_constraints}

\paragraph{X-rays.} A number of studies have investigated the properties of the
thermal gas in the Fermi bubbles and in the Galactic halo from X-ray
observations. The parameters can be either inferred from the thermal,
soft X-ray spectrum \citep{Kataoka:2013tma,Kataoka:2015dla} or from
individual Oxygen lines \citep{Miller:2016chr}. The gas densities
inferred are of the order $n_{\text{gas}} \sim 10^{-3} \,
\text{cm}^{-3}$ and the temperatures of the gas just outside the
bubbles vary between $kT \simeq 0.3 \, \text{keV}$ and $0.5 \,
\text{keV}$. This is higher than the canonical temperature of the
Galactic halo of $kT \simeq 0.2 \, \text{keV}$ and requires a heating
agent, perhaps a weak shock with a low Mach number; $\mathcal{M}
\simeq 1.5 \mathellipsis 2.3$. Finally, with the typical sound speed
in the Galactic halo of $c_s \simeq 200 \, \text{km} \,
\text{s}^{-1}$, one infers shock speeds of $v_{\rm sh} \simeq 300
\mathellipsis 500 \, \text{km} \, \text{s}^{-1}$.

The absence of evidence for a strong shock coinciding with the bubble
edge implies that diffusive shock acceleration at the bubble edge
cannot be the primary mechanism of acceleration. If electrons get
accelerated in the Galactic plane or even in a hypothetical
large-scale jet along the Galactic minor axis, they need to travel
over distances of several kpc without much energy loss to fill the
bubble volume. As a result they will suffer severe cooling losses and
a gradual softening of their spectrum, or even quench the electron
density completely. Note further that the low shock speeds found by
the X-ray modeling lead to even larger dynamical times than with the
$1000 \, \text{km} \, \text{s}^{-1}$ assumed above, making the energy
losses even more important.

\paragraph{Quasar absorption.} 

The observation of absorption by the gas associated with the bubbles
from a background quasar can also be used to set bounds on the outflow
speed. In the UV absorption lines from PDS 456 two (asymmetric)
components with velocities of $v \simeq -235$ and $+250 \, \text{km}
\, \text{s}^{-1}$ with respect to the local standard of rest could be
identified \citep{Miller:2016chr}. For the conical outflow assumed in
that study, this implies an upper limit on the outflow speed of
$\gtrsim 900 \, \text{km} \, \text{s}^{-1}$. This seems to be in
conflict with the shock speed inferred from the X-ray modelling
described above.  Note, however, that the outflow speed inferred from
the absorption lines of one quasar is very dependent on the assumed
geometry of the flow. Future observations of additional sight lines
towards other quasars can help mapping out the flow structure, thus
possibly also constraining it geometry, and might bring the results
into agreement with the values inferred from X-rays.

%------------------------------------------------------------------------------------------------------------------------------------------
%------------------------------------------------------------------------------------------------------------------------------------------
\subsection{Models}
\label{sec:models}

The Fermi bubbles have also generated a great deal of interest on the
modelling side
\citep{Crocker:2010dg,Cheng:2011xd,Cheng:2011tx,Zubovas:2011py,Mertsch:2011es,Zubovas:2012bn,Guo:2011eg,Yang:2012fy,Lacki:2013zsa,Crocker:2013mna,Fujita:2013jda,Thoudam:2013eaa,Yang:2013kca,Crocker:2013mna,Fujita:2014oda,Cheng:2014nva,Cheng:2014lca,Mou:2014pea,Crocker:2014fla,Cheng:2015zda,Mou:2015wxa,Sarkar:2015xta,Sasaki:2015nta,Yang:2017tjr}. The
variety of models is most conveniently classified by:
\begin{itemize}
\item the source of energy: super massive black hole or stellar winds/supernovae;
\item the acceleration region: jet or sources in the disk or \textit{in situ} (by shocks or turbulence);
\item the nature of the high-energy particles: hadrons or leptons.
\end{itemize}
Of course, the individual options are not mutually exclusive. For
instance, in hadronic models, the bulk of the high-energy gamma-rays
comes from decay of neutral pions. Charged pions, however, get
produced at similar rates and can, given the radiation fields, their
$e\pm$ byproducts can inverse-Compton scatter soft photons into low
energy gamma-rays.  However, in this particular scenario the
synchrotron spectrum would be too soft (Ackermann et al. 2014)

As a full discussion of all proposed models is beyond the scope of
this highlight presentation, only two particular classes of models
will be presented, and a few concrete examples will be shown.

\paragraph{Jet models.} Astrophysical jets are thought to be powered 
by accretion onto a spinning, compact object, like neutron stars or
black holes. Given the position and symmetry of the Fermi bubbles, the
supermassive black hole at the Galactic centre is a prime
candidate. Although conspicuously quiet (its X-ray luminosity is
currently more than 11 orders below the Eddington luminosity), there
is indirect evidence for earlier epochs of active accretion, e.g. from
X-ray reflections.

Jets are usually associated with high speeds $\gtrsim 1000 \,
\text{km} \, \text{s}^{-1}$. This allows for the electrons to be less
impacted by energy losses than in starburst/star formation models and
therefore the source of energisation of the high-energy electrons can
be in the Galactic disk or inside the jet. (Note, however, that the
jet speed is not necessarily directly implying the dynamical age as
the bubbles can be formed by a fountain-like back flow due to the
termination of the jet by the ram pressure of gas in the Galactic
halo.)

One of the earliest studies of a leptonic jet model employing a
hydrodynamical simulation \citep{Guo:2011eg} found that the
lateral extent of the Fermi bubbles could be explained if the jet was
underdense but slightly overpressured. If active at $10 \, \%$ of the
Eddington luminosity for $1 - 2 \, \text{Myr}$ until about a
$\text{Myr}$ ago, the morphology would match the observations. A
subsequent MHD simulation of the Fermi bubbles blown up by a jet
\citep{Yang:2012fy} showed further that the shock compression
at the bubble edges would compress the magnetic field such that it
gets aligned with the bubble edge. We will return to this point in
Sec.~\ref{sec:toroidal_coordinates}.

\paragraph{Star formation/star burst models.} The Galactic winds that get
collectively powered by an ensemble of stellar winds or supernova
activity, are usually operating at smaller speeds, $\lesssim 500 \,
\text{km} \, \text{s}^{-1}$. This implies a larger dynamic time-scales
than for the jet model, leading to a preference in the literature for
hadronic models, as leptons would cool too fast. In hadronic models,
on the other hand, cosmic rays need to be accumulated over much longer
time scales, given the low gas densities of the order of $10^{-3} \,
\text{cm}^{-3}$ (see Sec.~\ref{sec:other_constraints}), to produce the
observed gamma-ray fluxes. In turn, this and the observed hard
$E^{-2}$ spectrum require an effective confinement of the high-energy
cosmic rays to the bubbles and a suppression of (energy-dependent)
escape. (See, however, \citet{Keshet:2016fbq}.) The sources of high
energy particles are nevertheless oftentimes assumed to be in the
Galactic disk.

The most detailed numerical star formation/star burst model for the
Fermi bubbles as of yet \citep{Sarkar:2015xta} employs a
hydrodynamical code to investigate the interaction of a Galactic wind
with the circumgalactic medium. It is found that a luminosity of $5
\times 10^{40} \, \text{erg} \, \text{s}^{-1}$ and a density in the
halo of $10^{-3} \, \text{cm}^{-3}$ can reproduce the morphology
observed in gamma-rays and is also in agreement with X-ray
observations. Interestingly, this luminosity is close to the one
inferred from the current star formation rate, $\text{SFR} \simeq
0.007 \mathellipsis 0.1 M_{\odot} \, \text{yr}^{-1}$, when assuming an
efficiency of $30 \, \%$ for conversion into mechanical power,
$\mathcal{L} \simeq 10^{40} \text{erg} \, \text{s}^{-1}
\varepsilon_{0.3} (\text{SFR} / (0.1 M_{\odot} \, \text{yr}^{-1}))$.

The outflow from the inner Galaxy leads to a shock structure known
from the heliosphere or supernova remnants, with a radial forward
shock at $\sim 11 \, \text{kpc}$, a more tangled contact discontinuity
extending to $\sim 8 \, \text{kpc}$ above the Galactic centre and a
very much tangled reverse shock a few kiloparsecs inside of the
contact discontinuity. Thus, in this model, the edge of the gamma-ray
bubble does not coincide with the projection of the forward shock, but
rather the contact discontinuity. Whether this is due to the diffusion
prescription of \citet{Sarkar:2015xta} changing across the contact
discontinuity would need to be explored further.

%------------------------------------------------------------------------------------------------------------------------------------------
%------------------------------------------------------------------------------------------------------------------------------------------
\subsection{Motivation}

The observation of $\gamma$-rays from the bubbles implies a huge
reservoir of high-energy particles in the Galactic halo, but the
source and the mechanism of acceleration of these particles has not
been established thus far.  Other sources of non-thermal particles,
like supernova remnants, pulsar wind nebulae, jets in active galaxies
or winds in starburst galaxies, show evidence of shocks through X-rays
or ionization lines. The Fermi bubbles, however, show no such evidence
of a (strong) shock, raising the question of the possible mechanism of
acceleration. Acceleration by plasma turbulence (or ``stochastic
acceleration'', SA), however, can fill the bubbles with high-energy
electrons. (See, \citet{Petrosian:2012ba} for a recent review of SA.

A first SA model for the Fermi bubbles \citep{Mertsch:2011es} was
presented quickly after their discovery. This model was employing the
solution of a simplified version of the transport
equation. Specifically, diffusion was ignored as a spatial transport
process and advection was the only transport process. In this
framework, cosmic ray electrons are just passively advected with the
downstream flow while being stochastically accelerated.  The time
scale hierarchy $t_{\text{dyn}} \gg t_{\text{cool}} \gg
t_{\text{acc}}$ of dynamical, cooling and acceleration times, allows a
steady-state solution of the variation of electron spectrum with
radius for a given spatial variations of theses and and the escape
time, $t_{\text{esc}}$. While successful in explaining the overall
spectrum of the bubbles as well as the sharp edges, the lack of
diffusive transport was an important shortcoming. In addition, the
interstellar radiation fields on which the cosmic ray electrons
scatter was assumed homogeneous which must be an
oversimplification. What is needed is a detailed numerical model,
taking into account all the spatial transport processes (diffusion,
advection), energy losses (ionisation, bremsstrahlung, synchrotron,
inverse Compton scattering) and energy gains (shock and SA).

In the remainder of this paper, we will present our computation of the
SA of high-energy electrons in the Fermi bubbles.
Sec.~\ref{sec:method} introduces our method for solving the transport
equation on a grid that is suited for the geometry of the bubbles. We
will define three setups and specify the parameter values
considered. We will show our result for those three setups in
Sec.~\ref{sec:results} and comment on compatibility with observational
data. In Sec.~\ref{sec:summary_conclusions} we summarise and conclude.

%------------------------------------------------------------------------------------------------------------------------------------------
%------------------------------------------------------------------------------------------------------------------------------------------
%------------------------------------------------------------------------------------------------------------------------------------------
\section{Method}
\label{sec:method}

%------------------------------------------------------------------------------------------------------------------------------------------
%------------------------------------------------------------------------------------------------------------------------------------------
\subsection{Transport equation}

We start by considering the following transport equation for the
(isotropic) phase space density $f(\vec{r}, p, t)$,
e.g.~\citep{1987PhR...154....1B},
\begin{align}
\frac{\partial f}{\partial t} = \nabla \cdot \left( K \cdot \nabla f -
\vec{V} f \right) + \frac{1}{p^2} \dfrac{\partial}{\partial p} \left(p^2 D_{pp} \dfrac{\partial f}{\partial p} \right)
\nonumber
\\ + \frac{1}{p^2} \dfrac{\partial}{\partial p} p^2 \left( \frac{p}{3} \left( \nabla \cdot \vec{V}\right) f \right) \, . \label{eqn:transport_phasespacedensity}
\end{align}
Here, spatial transport is governed by the diffusion tensor $K$ and
the advection velocity $\vec{V}$, the latter also leading to adiabatic
gains/losses through the divergence term. Momentum space diffusion
depends on the diffusion coefficient $D_{pp}$.

For numerical convenience, we reformulate
eq.~\ref{eqn:transport_phasespacedensity} in terms of $\psi=4 \pi
p^2\times f$, the differential (in momentum) particle density which is related
to the total particle density $n = \int \mathrm{d} p \, \psi$.
We also add momentum and  catastrophic losses 
$-\partial (\dot{p} \psi) / \partial p$
and $-\psi / \tau$ and a source term $S$,
\begin{align}
\frac{\partial \psi}{\partial t} = \nabla \cdot \left( K \cdot \nabla \psi
- \vec{V} \psi \right) + \dfrac{\partial}{\partial p} \left( p^2 D_{pp} \dfrac{\partial}{\partial p} \frac{\psi}{p^2}
\right) \nonumber
\\ + \dfrac{\partial}{\partial p} \left( - \dot{p} \psi + \frac{p}{3} \left( \nabla \cdot \vec{V}\right) \psi \right) - \frac{\psi}{\tau} + S \, .
\label{eqn:transport}
\end{align}

%------------------------------------------------------------------------------------------------------------------------------------------
%------------------------------------------------------------------------------------------------------------------------------------------
\subsection{Shock equation}

At the shock, we need to carefully evaluate the transport
eq.~\ref{eqn:transport} because of the discontinuity in $\vec{V}$ (and
in other transport parameters). We denote quantities upstream
(downstream) of the shock by a minus (plus) sign. We demand $\psi$ to
be continuous across the shock,
\begin{equation*}
\psi^- = \psi^+ \, ,
\end{equation*}
and allow for the presence of sources at the shock, $S^*
\delta(\vec{r} - \vec{r}_{\text{sh}})$, so that by continuity
\begin{equation*}
\nabla \cdot \vec{J} = S^* \delta(\vec{r} - \vec{r}_{\text{sh}}) \, .
\end{equation*}
With Gauss' theorem, we can write this as
\begin{equation}
\int_V \mathrm{d} V \, (\nabla \cdot \vec{J}) = \int_A \mathrm{d} \vec{A} \cdot \vec{J} = \int_V \mathrm{d} V \, S^* \delta(\vec{r} - \vec{r}_{\text{sh}}) \, .
\label{eqn:shock}
\end{equation}
The \emph{particle density} flux $\vec{J}$ is here
\begin{align}
\vec{J} &= - \hat{n} K_{\parallel} (\hat{n} \cdot \nabla) \psi + \frac{1}{3} \left(2 - \frac{\partial \ln \psi}{\partial \ln p} \right) \vec{V} \psi \, . \label{eqn:particle_density_flux}
\end{align}

For numerical solution of the transport and shock equations, we need
to specify a coordinate system.

%------------------------------------------------------------------------------------------------------------------------------------------
%------------------------------------------------------------------------------------------------------------------------------------------
\subsection{Coordinates}
\label{sec:toroidal_coordinates}

A convenient choice of coordinates should help simplify the
computation, e.g.\ in that it eases or altogether eliminates the
transformation from the simulation coordinates to the frame in which
the diffusion tensor is diagonal. The method for treating the
discontinuity of the shock requires that the shock normal to be
aligned with one coordinate direction.

\begin{figure}[bt]
\centering
\includegraphics[width=0.5\textwidth]{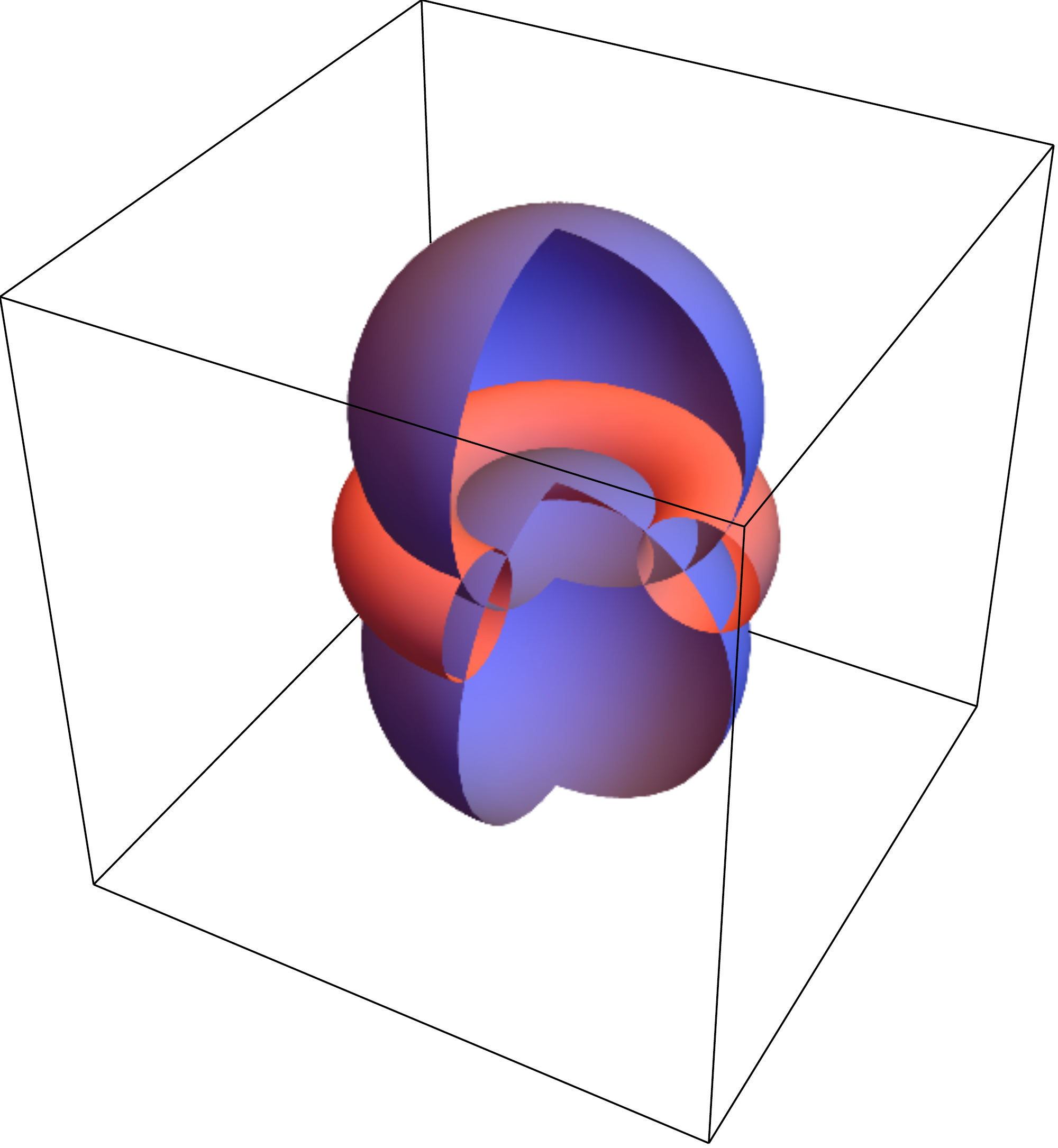}
\caption{Toroidal coordinates. Surfaces of constant $v$ are tori of
  varying radii surrounding the foci ring of radius $a$ and are shown
  in red. Surfaces of constant $u$ are spheres of varying radii that
  all intersect the foci ring and are shown in blue.}
\label{fig:surfaces}
\end{figure}

Given the bi--lobular shape of the bubbles as observed in
gamma--rays~\citep{Su:2010qj,Fermi-LAT:2014sfa} and used in the (M)HD
simulations~\citep{Guo:2011eg,Yang:2012fy}, leads us to employ
toroidal coordinates ($u, v, \phi$) which map to cartesian coordinates
($x, y, z$) through
\begin{align}
x &= \frac{a \sinh{v} \cos{\phi}}{\cosh{v} - \cos{u}} \, , \label{eqn:x_uvphi} \\
y &= \frac{a \sinh{v} \sin{\phi}}{\cosh{v} - \cos{u}} \, , \label{eqn:y_uvphi} \\
z &= \frac{a \sin{u}}{\cosh{v} - \cos{u}} \, . \label{eqn:z_uvphi}
\end{align}
In Fig.~\ref{fig:surfaces}, we show a plot of surfaces of constant $u$
and $v$. Surfaces of constant $u$ are spheres of varying radii that
all intersect a foci ring of radius $a$. Surfaces of constant $v$ are
tori of varying radii surrounding the foci ring. We set $a = 1 \,
\text{kpc}$ throughout unless otherwise noted.

From the transformation between cartesian and toroidal coordinates,
eqs.~\ref{eqn:x_uvphi} - \ref{eqn:z_uvphi}, we can compute the scale
factors
\begin{align}
h_u &= h_v = \frac{a}{\cosh v - \cos u} \, \quad \text{and} \quad h_{\phi} &= \frac{a \sinh v}{\cosh v - \cos u} \, .
\end{align}

Here and in the following, we assume azimuthal ($\phi$) symmetry and
also define the cylindrical radial coordinate $\rho = \sqrt{x^2 +
  y^2}$.

%------------------------------------------------------------------------------------------------------------------------------------------
\subsubsection{Transport equation in toroidal coordinates}

We can write the transport equation~\ref{eqn:transport} in toroidal
coordinates,
\begin{align}
\frac{\partial \psi}{\partial t} &= \left[ \frac{(\cosh v - \cos u)^2}{a^2} K_{uu} \right] \frac{\partial^2 \psi}{\partial u^2} \nonumber
\\ & + \left[ - \frac{(\cosh v - \cos u) \sin u}{a^2} K_{uu} + \frac{(\cosh v - \cos u)^2}{a^2} \dfrac{\partial K_{uu}}{\partial u} \right] \frac{\partial \psi}{\partial u} \nonumber
\\ & + \left[ - \frac{\cosh v - \cos u}{a} V_u \right] \frac{\partial \psi}{\partial u} \nonumber
\\ & + \left[ \frac{(\cosh v - \cos u)^2}{a^2} K_{vv} \right] \frac{\partial^2 \psi}{\partial v^2} \nonumber
\\ & + \left[ \frac{(\cosh v - \cos u)(1 - \cosh v \cos u)}{a^2 \sinh v} K_{vv} \right. \nonumber
\\ & \quad\quad \left. + \frac{(\cosh v - \cos u)^2}{a^2} \dfrac{\partial K_{vv}}{\partial v} \right] \frac{\partial \psi}{\partial v} \nonumber
\\ & + D_{pp} \dfrac{\partial^2 \psi}{\partial p^2} + \left[ \dfrac{\partial D_{pp}}{\partial p} - \frac{2}{p} D_{pp} \right] \dfrac{\partial \psi}{\partial p} \nonumber
\\ & + \left[ -\dot{p} + \frac{p}{3} (\nabla \cdot \vec{V} ) \right] \dfrac{\partial \psi}{\partial p} \nonumber
\\ & + \left[ \frac{2}{p^2} D_{pp} - \frac{2}{p} \dfrac{\partial D_{pp}}{\partial p} - \dfrac{\partial \dot{p}}{\partial p} - \frac{2}{3} (\nabla \cdot \vec{V}) - \frac{1}{\tau} \right] \psi + S \, . \label{eqn:transport_toroidal}
\end{align}

A comment about the velocity divergence,
\begin{equation*}
\nabla \cdot \vec{V} = \frac{(\cosh v - \cos u)^3}{a^3 \sinh v} \dfrac{\partial}{\partial u} \left( \frac{a^2 \sinh v}{(\cosh v - \cos u)^2} V_u \right) \, ,
\end{equation*}
is in order. For an incompressible flow, $0 \equiv \nabla \cdot
\vec{V}$, but if we want fronts to follow lines of constant $u$, then
we require the $v$--dependence (see, discussion in
Sec.~\ref{sec:setup1_parameters} below)
\begin{equation*}
V_u \propto h_u = \frac{a}{\cosh v - \cos u} \, .
\end{equation*}
Note that in eq.~\ref{eqn:transport_toroidal}
\begin{equation}
\frac{(\cosh v - \cos u)(1 - \cosh v \cos u)}{a^2 \sinh v} \frac{\partial \psi}{\partial v}
\end{equation}
is indeterminate for $v \to 0$ as $\sinh v \to 0$ and $\partial \psi /
\partial v \to 0$ (due to symmetry). Employing the L'Hopital's rule,
we can replace this by
\begin{equation}
\frac{(\cosh v - \cos u)(1 - \cosh v \cos u)}{a^2 \cosh v} \frac{\partial^2 \psi}{\partial v^2}
\end{equation}

%------------------------------------------------------------------------------------------------------------------------------------------
\subsubsection{Shock equation in toroidal coordinates}

If the shock is in the $u$--plane and the flux is perpendicular to the
shock, $\vec{J} = J_u \hat{u}$, eq.~\ref{eqn:shock} reads
\begin{equation}
\quad J_u^+ - J_u^- = \lim_{\epsilon \to 0} \int_{u_{\text{sh}} - \epsilon}^{u_{\text{sh}} + \epsilon} \mathrm{d} u \, h_u S^* \delta(u - u_{\text{sh}}) = h_u S^* \, ,
\end{equation}
where (cf. eq.~\ref{eqn:particle_density_flux})
\begin{equation}
J_u = -K_{uu} \frac{1}{h_u} \dfrac{\partial \psi}{\partial u} + \frac{1}{3} \left(2 - \frac{\partial \ln \psi}{\partial \ln p} \right) V_u \psi \, .
\end{equation}
We thus find,
\begin{align}
- \left( \frac{1}{h_u} V_u^- - \frac{1}{h_u} V_u^+ \right) \frac{1}{3} \left(2 \psi - p \dfrac{\partial \psi}{\partial p} \right) \nonumber
\\ + \left( \frac{K_{uu}^-}{h_u^2} \left( \dfrac{\partial \psi}{\partial u} \right)^- - \frac{K_{uu}^+}{h_u^2} \left( \dfrac{\partial \psi}{\partial u} \right)^+ \right) &= S^*(p) \, , \label{eqn:shock_toroidal}
\end{align}

%------------------------------------------------------------------------------------------------------------------------------------------
%------------------------------------------------------------------------------------------------------------------------------------------
\subsection{Finite--difference method}

Parabolic partial differential equations like the transport
equation~\ref{eqn:transport} are oftentimes solved numerically by
finite difference methods. Here, we numerically solve the transport
equation in toroidal coordinates
(cf. eq.~\ref{eqn:transport_toroidal}) in the widely used
Crank--Nicolson scheme~\citep{1947PCPS...43...50C}. This is a
semi--implicit method which results in a tridiagonal system that can
be efficiently solved by the Thomas algorithm. The difficulty in the
case at hand is the presence of the shock which breaks the
tridiagonality of the involved matrix. In particular at the shock
position, we solve the shock equation~\ref{eqn:shock_toroidal} instead
of the transport equation~\ref{eqn:transport_toroidal}. Here we follow
a method outlined by \citet{Langner:2004zz} which treats transport in
the heliosphere in the presence of the helioshperic termination shock.

%------------------------------------------------------------------------------------------------------------------------------------------
%------------------------------------------------------------------------------------------------------------------------------------------
\subsection{Computational grid}

The computational grid is three--dimensional: two spatial, toroidal
coordinates ($u$ and $v$) and one momentum coordinate ($p$). Choosing
the spatial grids to be linear renders the coefficients for the finite
difference scheme particularly simple and offers the added advantage
of fine resolution close the Galactic centre and at the bases of the
bubbles. For the momentum grid we chose logarithmic spacing in order
to evenly sample the spectra which will be close to power law:
\begin{align}
u_i &= u_\mathrm{min} + i \Delta r = u_\mathrm{min} + \frac{i}{n} (u_\mathrm{max} - u_\mathrm{min}) \, , \quad i = 0, \mathellipsis n, \\
v_j &= v_\mathrm{min} + j \Delta \theta = v_\mathrm{min} + \frac{j}{m} (v_\mathrm{max} - v_\mathrm{min}) \, , \quad j = 0, \mathellipsis m, \\
p_k &= p_\mathrm{min} \mathrm{e}^{k \Delta \ln p} = p_\mathrm{min} \mathrm{e}^{k \left[ \ln \left( p_\mathrm{max} / p_\mathrm{min} \right) \right] / q} \, , \quad k = 0, \mathellipsis q.
\end{align}

For the minimum and maximum coordinate values and number of grid
points, we need to balance accuracy and computational speed under the
constraint of suppressing numerical artefacts,
e.g.\ oscillations. Here, we have chosen the following grid
parameters:
\begin{equation}
	\begin{array}{rlrlrl}
		u_{\text{min}} &= 0 \, ,					& u_{\text{max}} &= \pi \, ,					& n &= 800 \, , \\
		v_{\text{min}} &= 0 \, ,					& v_{\text{max}} &= 2 \, ,					& m &= 40 \, , \\
		c p_{\text{min}} &= 10^{-3} \, \text{GeV} \, ,	& c
p_{\text{max}} &= 10^{4} \, \text{GeV} \, ,	& q &= 140 \, . \\
	\end{array}
\end{equation}
For $u_{\text{max}} = \pi$ and $v_{\text{max}} \to \infty$, the
computational domain is covering the whole $\rho$--$z$ plane. To limit
the size of the grid while assuming linear spacing, we limit
$v_{\text{max}}$ to finite values. This will affect the transport and
acceleration of particles close to $\rho = 1 \, \text{kpc}$; however,
due to the presence of strong diffuse, conventional emission, the
Galactic disk is usually excluded from diffuse studies, cf.,
e.g.,~\citep{Su:2010qj,Fermi-LAT:2014sfa}.

\begin{figure}
\centering
\includegraphics[width=0.45\textwidth]{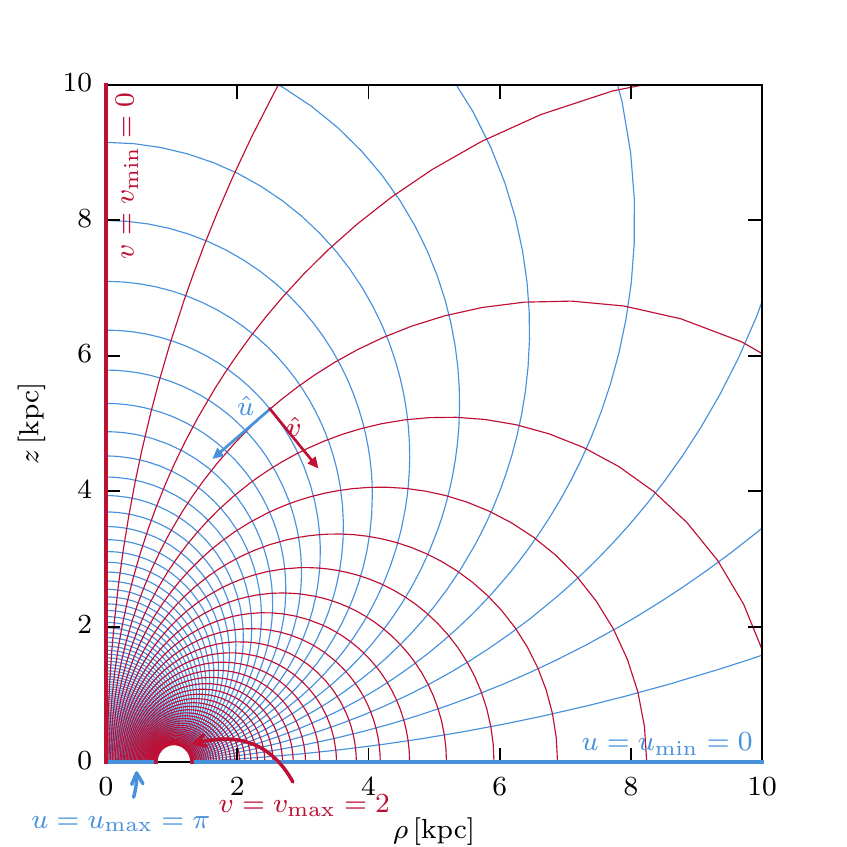}
\caption{The spatial part of the computational grid. Note that for
  clarity, we only show $n=100$ spacings in the $u$--direction here,
  whereas in the numerical simulations, we have chosen $n = 800$
  throughout. Note also that the use of $v_{\rm max}=2$ instead of
  $\infty$ ignores a small region of space.}
\label{fig:grid}
\end{figure}

The spatial part of the computational grid is shown in
Fig.~\ref{fig:grid}.

%------------------------------------------------------------------------------------------------------------------------------------------
%------------------------------------------------------------------------------------------------------------------------------------------
\subsection{Parameters}

\begin{figure*}[tbh]
\includegraphics[scale=1]{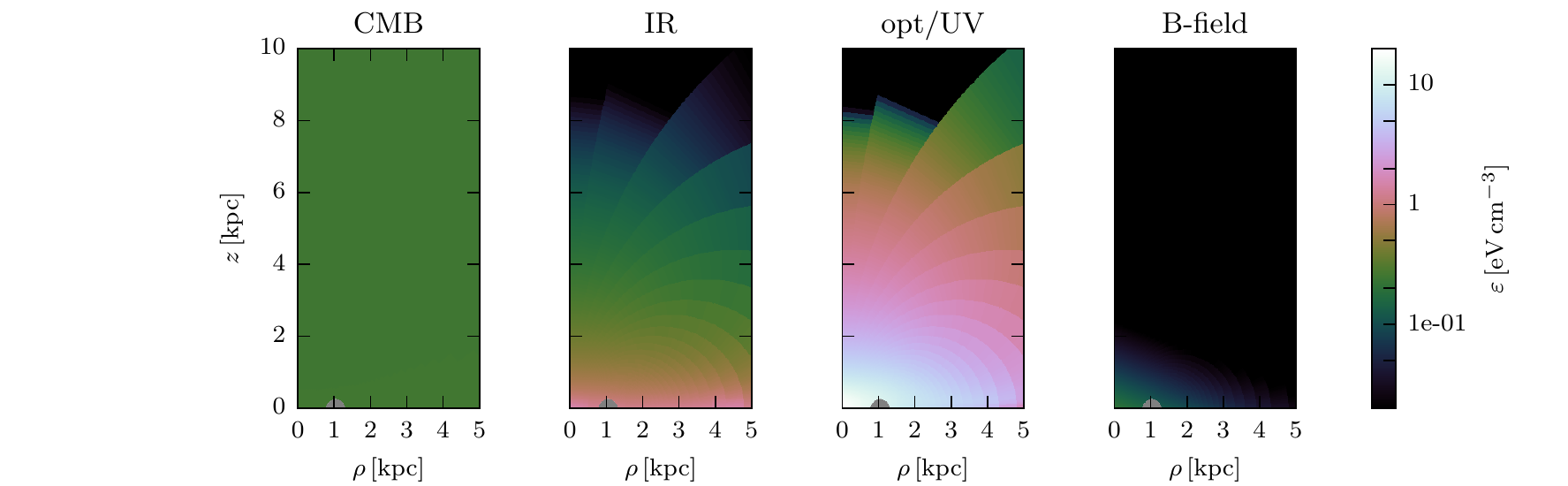}
\caption{Energy densities $\varepsilon$ of the CMB, IR and optical/UV
  parts of the ISRF as well as the energy density of the $B$--field in
  the $\rho$--$z$--plane. We define the energy ranges as $4.3 \times
  10^{-5}$ to $1.4 \times 10^{-2} \, \text{eV}$ for the CMB, $1.4
  \times 10^{-2}$ to $0.22 \, \text{eV}$ for the IR and $0.22$ to $14
  \, \text{eV}$ for the optical/UV.}
\label{fig:setup1_isrfs}
\end{figure*}

Diffusion is only really isotropic in the limit of a small regular
magnetic field $B_0$, i.e. when the fractional turbulence level $\eta=
\delta B^2 / (B_0^2 + \delta B^2) \to 1$. In the general case, the
symmetric part of the diffusion tensor $K$ can be written as $K =
\text{diag}(K_{\perp}, K_{\perp}, K_{\parallel})$ in cartesian
coordinates where without loss of generality we have assumed that
$\vec{B}_0 \parallel \hat{z}$. In quasi--linear theory,
$K_{\parallel}$ and $K_{\perp}$ scale differently with the turbulence
level $\eta)$; $K_{\parallel} \propto \eta^{-1}$ while $K_{\perp}
\propto \eta$. In our case, we assume $\vec{B}_0 \parallel \hat{v}$
and so $K_{\perp} = K_{uu} \propto \eta$ and $K_{\parallel} = K_{vv}
\propto \eta^{-1}$, and adopt a momentum dependence consistent with
resonant interactions with Kolmogorov turbulence,
\begin{align}
K = \left( \begin{array}{cc} K_{uu} & 0 \\ 0 & K_{vv} \end{array} \right) \quad \text{with} \quad K_{uu} &= \beta K_{uu,0} \left( \frac{pc}{1 \, \text{GeV}} \right)^\delta
\\ \text{and} \quad K_{vv} &= \beta K_{vv,0} \left( \frac{pc}{1 \, \text{GeV}} \right)^\delta \, .
\end{align}

For the momentum diffusion coefficient, we are employing a relation
from quasi--linear theory,
\begin{equation}
K_{\parallel} D_{pp} \approx p^2 V_A^2 \, ,
\end{equation}
where $V_A$ is the Alfv\'en speed. This fixes the momentum dependence
of $D_{pp}$,
\begin{equation}
D_{pp} = \beta^{-1} D_{pp,0} \left( \frac{pc}{1 \, \text{GeV}} \right)^{2-\delta} \, ,
\end{equation}
and parametrises its normalisation relative to $K_{\parallel}$ through
the Alfv\'en speed $V_A \simeq B_0 / \sqrt{4 \pi m_p n_{\text{gas}}}$,
\begin{align}
D_{pp,0} =& \frac{V_A^2}{K_{\parallel}} \left( \frac{\text{GeV}}{c} \right)^2
\\ \simeq& \left( \frac{V_A}{300 \, \text{km} \, \text{s}^{-1}} \right)^2 \left( \frac{K_{\parallel}}{3 \times 10^{28} \, \text{cm}^2 \, \text{s}^{-1}} \right)^{-1} 
\\ &\quad \times \left( \frac{\text{GeV}}{c} \right)^2 \text{Myr}^{-1} \, .
\end{align}

We assume the shock to follow an isocontour in $u$ and in order for
advection fronts to follow lines of constant $u$, we choose the
advection velocity $\vec{V} = V_u \hat{u}$ and its absolute value
proportional to the scale factor $h_u = a/(\cosh v - \cos u)$. We
introduce an additional $u$--dependence, $(1 - \cos u)$, and define
the compression ratio $r$ of the shock, that is the ratio of upstream
to downstream speed in the shock frame, such that
\begin{equation}
V_u = - \left(1 - \frac{1}{r} \right) V_{\text{sh}} \frac{a (1 - \cos u)}{\cosh v - \cos u} \, ,
\label{eqn:Vu}
\end{equation}
This results in the shock travelling with constant speed
$V_{\text{sh}}$ along the $z$--axis:
\begin{align*}
\frac{\mathrm{d} u}{\mathrm{d} t} = \frac{\cosh v - \cos u}{a} V_u &= - V_{\text{sh}} (1 - \cos u) \, , \\
\Leftrightarrow -\cot(u/2) + \cot(u_0/2) &= - V_{\text{sh}} (t - t_0) \, .
\end{align*}
With $t_0 = 0$ and $u_0 = u(t_0) = \pi$,
\begin{equation*}
u_{\text{sh}}(t) = 2 \, \mathrm{arccot} (V_{\text{sh}}t) \, ,
\end{equation*}
and one finds
\begin{equation*}
z(u_{\text{sh}}(t), 0) = \frac{a \sin u_{\text{sh}}(t)}{1 - \cos u_{\text{sh}}(t)} = \frac{a \sin (2 \, \mathrm{arccot} V_{\text{sh}} t)}{1 - \cos (2 \, \mathrm{arccot} V_{\text{sh}} t)} = V_{\text{sh}} t \, .
\end{equation*}
With the $V_{\text{sh}} = 3 \times 10^{-7} \, \text{kpc} \, \text{yr}
\simeq 300 \, \text{km} \, \text{s}^{-1}$ that we adopt, the shock
reaches a height of $z = 6 \, \text{kpc}$ in $t = 20 \, \text{Myr}$.
And with the  $u$--dependence in eq.~( \ref{eqn:Vu}) we get
\begin{equation*}
\nabla \cdot \vec{V} = \left(1 - \frac{1}{r} \right) V_{\text{sh}} \sin u \left( 1 - 3 \frac{(1 - \cos u)}{\cosh v - \cos u} \right) \, .
\end{equation*}
Here, we set the compression ratio to $4$, but it is strictly only so
along the $z$--axis.

One can estimate the timescales for shock acceleration and stochastic
acceleration as $\tau_{\rm sh} \sim K_{uu} / V_{\text{sh}}^2$ and
$\tau_{\rm SA} \sim p^2 / D_{pp} \sim K_{vv} / V_A^2$ with the ratio
is $(\tau_{\rm sh}/\tau_{\rm SA}) \sim (K_{uu} / K_{vv}) V_A^2 /
V_{\text{sh}}^2=(\eta/{\cal M}_A)^2$, where ${\cal M}_A$ is the
Alf\'ven Mach number. (see, Petrosian 2012). In the following, we
adopt $V_A = 300 \, \text{km} \text{s}^{-1}$ everywhere. The shock
velocity, however, is not constant along the shock as the bubble
expands more slowly laterally than vertically. At the top,
$v_{\text{sh}} = 300 \, \text{km} \text{s}^{-1}$ which coincides with
the Alfv\'en speed and thus the rates for shock acceleration and
stochastic acceleration will be equal for isotropic diffusion ($K_{uu}
= K_{vv}$). For anisotropic diffusion ($K_{uu} < K_{vv}$), shock
acceleration even operates faster than stochastic acceleration. In
contrast, at the foot of the bubble, the shock velocity is much lower
than the Alfv\'en speed, rendering shock acceleration
inefficient. Note that in either case, the effective volume where
shock acceleration operates is rather small because of the small
diffusion length $K_{uu} / V_u$.

We idealise the shock as a surface of constant pseudo--radius
$u$. Furthermore, we assume that the (large--scale) $B$--field (which
defines the coordinate system in which the diffusion tensor is
diagonal) is aligned with these surface of constant $u$, that is
$\vec{B} \parallel \hat{v}$ where $\hat{v}$ is the unit vector of the
pseudo--polar coordinate, $v$.

The interstellar radiation fields (ISRFs) affect both the momentum
losses of the electrons and the generation of gamma--rays through
inverse--Compton scattering~\citep{Blumenthal:1970gc}. Here we adopt a
model\footnote{\url{http://galprop.stanford.edu/FITS/MilkyWay\_DR0.5\_DZ0.1\_DPHI10\_RMAX20\_ZMAX5\_galprop\_format.fits}}~\citep{Porter:2005qx}
from version 50 of the \texttt{GALPROP}
code~\citep{Moskalenko:1997gh,Orlando:2013ysa}. This contains
the energy density of the ISRF on a grid in cylindrical
coordinates. We bilinearly interpolate from the cylindrical grid to
our toroidal grid. As the cylindrical grid only extends to $z = \pm 5
\, \text{kpc}$, we linearly extrapolate for $|z| > 5 \, \text{kpc}$,
but set the ISRF to zero if it were otherwise negative. The left three
panels of Fig.~\ref{fig:setup1_isrfs} show the energy densities in the
CMB, IR and UV/optical ranges as a function of $\rho$ and $z$.

The coherent magnetic field is assumed to follow lines of constant $u$
(see above), that is $\vec{B} \parallel \hat{v}$, but we only use this
to define the coordinates in which the diffusion tensor is
diagonal. For synchrotron losses and the computation of
radio/microwave fluxes we ignore the regular field for the time
being. Instead, we only consider a turbulent component with rms value
\begin{equation}
B_{\text{rms}}(\rho, z) = B_0 \exp \left[ -\frac{\rho}{\rho_0} - \frac{z}{z_0} \right] \, .
\end{equation}
Here and in the following, we choose the set of parameters $B_0 = 3\,
\mu\text{G}$, $\rho_0 = 5 \, \text{kpc}$ and $z_0 = 1 \, \text{kpc}$
as a fiducial model. The energy density of this turbulent magnetic
field is shown in the rightmost panel of Fig.~\ref{fig:setup1_isrfs}.

For the source term $Q$, we simply adopt a Dirac delta function, both
in position and in momentum,
\begin{equation}
Q(\rho, z, p) \propto \delta(\vec{r}) \delta(p) \, .
\end{equation}
The normalisation is determined by fitting to the gamma--ray data from
\textit{Fermi}--LAT~\citep{Fermi-LAT:2014sfa}. Specifically, we
require the maximum gamma-ray flux in our map at $10 \, \text{GeV}$ to
be $E^2 J = 8 \times 10^{-7} \mathrm{GeV} \, \mathrm{cm}^{-2} \,
\mathrm{s}^{-1} \, \mathrm{sr}^{-1}$.

In principle we would have liked to also investigate the possibility
of accelerating electrons from the thermal background. However, for an
ambient temperature $T$ this would have required extending the
momentum grid down to thermal momenta of the order $p c \simeq k_B T
\simeq 8.6 \, \text{keV} \, (T/(10^8 \, \text{K}))$, that is by an
additional three orders of magnitude. Apart from increasing the size
of the momentum grid by more than $40 \, \%$, the short acceleration
time $t_{\text{sa}} = p^2/D_{pp} \propto p^\delta$ at low momenta
would have required much finer time--stepping (by a factor $\sim 10$),
significantly increasing the computational cost further.

\begin{figure*}[tbh]
\centering
\includegraphics[scale=1]{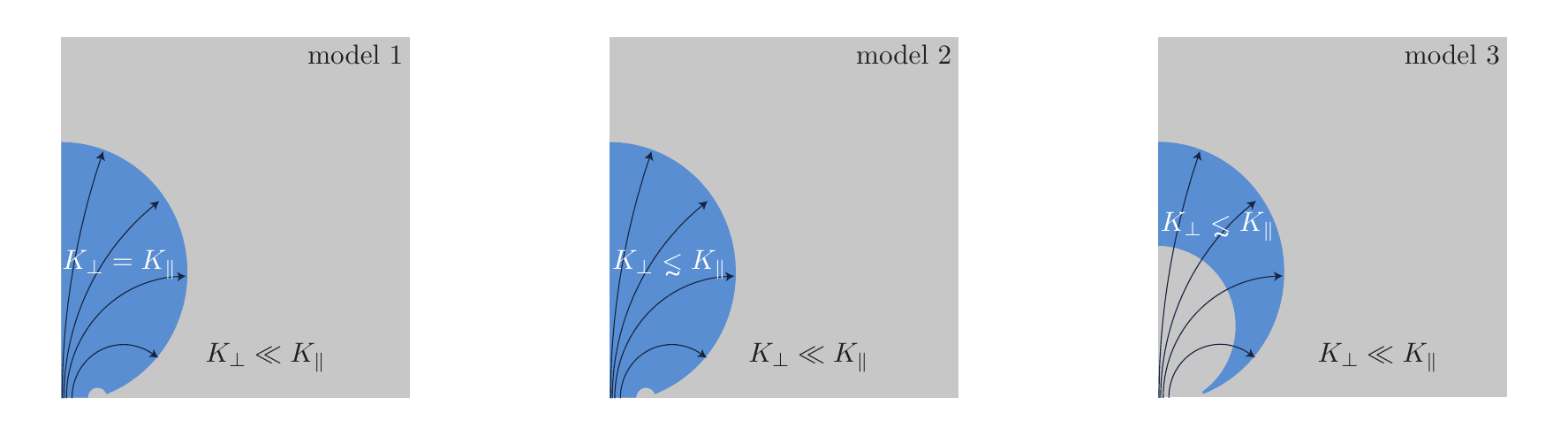}
\caption{Schematic overview of the three models. (See text for explanations.)}
\label{fig:models}
\end{figure*}

In the following we present three exemplary setups for the bubbles,
showing a conceptual evolution from the simplest possible model that
however fails, to a more complicated model that can reproduce the data
sufficiently well. For each setup, we detail and justify our parameter
choices before comparing our results to the available gamma--ray and
microwave data. See Fig.~\ref{fig:models} for a schematic overview of
the three setups, but refer to the text below for explanations.

\begin{table*}[tbh]
\centering
\caption{Summary of parameter choices in models 1, 2 and 3.}
\label{tbl1}
\begin{tabular}{l | c c | c c | c c c | l}
\hline
parameter										& \multicolumn{2}{c |}{model 1}	& \multicolumn{2}{c |}{model 2}	& \multicolumn{3}{c |}{model 3}	& comment		\\
											& inside		& halo		& inside	& halo			& inside	& shell	& halo	&				\\
\hline
$K_{uu,0} \, [10^{-7} \, \text{kpc}^2 \, \text{yr}^{-1}]$		& 1		& 0.1				& 1			& 0.1			& 0.1		& 1		& 0.1		& perpendicular diffusion coefficient	\\
$K_{vv,0} \, [10^{-7}	\, \text{kpc}^2 \, \text{yr}^{-1}]$		& 1		& 10				& 10			& 100		& 100	& 10		& 100	& parallel diffusion coefficient		\\
$D_{pp,0} \, [10^{-7} \, (\text{GeV}/c)^2 \, \text{yr}^{-1}]$	& 10		& 1				& 10			& 1			& 1		& 10		& 1		& momentum diffusion coefficient	\\
$\delta$										& \multicolumn{2}{c |}{$1/3$}	& \multicolumn{2}{c |}{$1/3$}	& \multicolumn{3}{c |}{$1/3$}	& spectral index				\\
$V_{\text{sh}} \, [10^{-7} \, \text{kpc} \, \text{yr}^{-1}]$		& \multicolumn{2}{c |}{$3$}	& \multicolumn{2}{c |}{$3$}	& \multicolumn{3}{c |}{$3$}	& shock speed					\\
$r$											& \multicolumn{2}{c |}{$4$}	& \multicolumn{2}{c |}{$4$}	& \multicolumn{3}{c |}{$4$}	& compression ratio				\\
\hline
\end{tabular}
\end{table*}

%------------------------------------------------------------------------------------------------------------------------------------------
\subsubsection{Model 1: isotropic diffusion}
\label{sec:setup1_parameters}

In the first setup, we consider diffusion \emph{inside} the bubbles to
be isotropic,
\begin{equation}
K_{uu} = K_{vv} = \beta K_0 \left( \frac{pc}{1 \, \text{GeV}} \right)^\delta \, ,
\label{eqn:isotropic_diffusion_coefficient}
\end{equation}
where $\beta = v_{\text{particle}}/c$, $K_{0,uu} = K_{0,vv} = 10^{-7}
\, \text{kpc}^2 \, \text{yr}^{-1} \approx 3 \times 10^{28} \,
\text{cm}^2 \, \text{s}^{-1}$ and $\delta = 1/3$. This is close to the
diffusion coefficient inferred from the boron--to--carbon ratio
measured at the solar position, $K_{\text{iso}}(1 \, \text{GeV})
\simeq 4.1 \times 10^{28} \, \text{cm}^2 \,
\text{s}^{-1}$~\citep{Trotta:2010mx}. \emph{Outside} the
bubbles, in the Galactic halo, we adopt $K_{0,uu} = 10^{-8} \,
\text{kpc}^2 \, \text{yr}^{-1} \approx 3 \times 10^{27} \, \text{cm}^2
\, \text{s}^{-1}$ and $K_{0,vv} = 10^{-6} \, \text{kpc}^2 \,
\text{yr}^{-1} \approx 3 \times 10^{29} \, \text{cm}^2 \,
\text{s}^{-1}$, that is diffusion is markedly anisotropic $(\eta =
10)$, with particles diffusing faster along the $v$--direction than
along the $u$--direction.

As a fiducial value for the momentum diffusion coefficient, we here
adopt $D_{pp,0} = 10^{-6} (\text{GeV}/c)^2 \, \text{yr}^{-1}$
\emph{inside} the bubbles, so an acceleration time
\begin{equation}
t_{\text{sa}} = p^2/D_{pp} \simeq 10^6 \, \text{yr} \, (pc / 1 \, \text{GeV})^{\delta} \, . \label{eqn:t_sa}
\end{equation}
The Alfv\'en speed $V_A \simeq 300 \, \text{km} \, \text{s}^{-1}$ can
be accommodated by, e.g., $B_0 \simeq 6 \, \mu\text{G}$ and
$n_{\text{gas}} \simeq 2 \times 10^{-3} \,
\text{cm}^{-3}$. \emph{Outside} the bubbles, we set $D_{pp,0} =
10^{-7} (\text{GeV}/c)^2 \, \text{yr}^{-1}$, in line with the usual
scaling.

%------------------------------------------------------------------------------------------------------------------------------------------
%------------------------------------------------------------------------------------------------------------------------------------------
\subsubsection{Model 2: anisotropic diffusion}

For the second setup, we consider the possibility that also diffusion
\emph{inside} the bubbles is anisotropic. Specifically, we adopt
$K_{uu,0} = K_{\parallel,0} = 10^{-7} \, \text{kpc}^2 \,
\text{yr}^{-1}$ and $K_{vv,0} = K_{\parallel,0} = 10^{-6} \,
\text{kpc}^2 \, \text{yr}^{-1}$ \emph{inside} and $K_{uu,0} =
K_{\parallel,0} = 10^{-8} \, \text{kpc}^2 \, \text{yr}^{-1}$ and
$K_{vv,0} = K_{\parallel,0} = 10^{-5} \, \text{kpc}^2 \,
\text{yr}^{-1}$ \emph{outside} the bubbles. We keep $\delta$ at $1/3$.

We set $D_{pp,0} = 10^{-6} (\text{GeV}/c)^2 \, \text{yr}^{-1}$
\emph{inside} the bubbles and $D_{pp,0} = 10^{-7} (\text{GeV}/c)^2 \,
\text{yr}^{-1}$ \emph{outside} the bubbles. All the other parameters
plus the ISRFs and the $B$--field are as in the first model,
cf.\ Sec.~\ref{sec:setup1_parameters}.

%------------------------------------------------------------------------------------------------------------------------------------------
%------------------------------------------------------------------------------------------------------------------------------------------
\subsubsection{Model 3: anisotropic diffusion and turbulent shell}

In the presence of a source of turbulence, the assumption of (almost)
isotropic diffusion of the first (second) setup can be justified. In
the following we assume that the shock itself is generating such
turbulence through hydrodynamic (e.g.\ Raleigh--Taylor or
Kelvin--Helmholtz) instabilities. To take into account that this
turbulence could be dissipated at large, kiloparsec distances from the
shock, we constrain this region to a shell behind the shock and assume
strongly anisotropic diffusion in the rest of the bubble volume
(c.f.\ the right panel of Fig.~\ref{fig:models}). In particular, we
choose
\begin{align}
K_{uu,0}^{\text{inside}} &= 10^{-8} \, \text{kpc}^2 \, \text{yr}^{-1} \quad \text{and} \quad K_{vv,0}^{\text{inside}} = 10^{-5} \, \text{kpc}^2 \, \text{yr}^{-1} \\
K_{uu,0}^{\text{shell}} &= 10^{-7} \, \text{kpc}^2 \, \text{yr}^{-1} \quad \text{and} \quad K_{vv,0}^{\text{shell}} = 10^{-6} \, \text{kpc}^2 \, \text{yr}^{-1} \\
K_{uu,0}^{\text{halo}} &= 10^{-8} \, \text{kpc}^2 \, \text{yr}^{-1} \quad \text{and} \quad K_{vv,0}^{\text{halo}} = 10^{-5} \, \text{kpc}^2 \, \text{yr}^{-1} \, .
\end{align}

This setup has the added benefit that according to quasi--linear
theory, the stochastic acceleration rate $D_{pp}/p^2$ is also enhanced
in a thin shell,
\begin{align}
D_{pp,0}^{\text{inside}} &= \times 10^{-7} (\text{GeV}/c)^2 \, \text{yr}^{-1} \, , \\
D_{pp,0}^{\text{shell}} &= \times 10^{-6} (\text{GeV}/c)^2 \, \text{yr}^{-1} \, , \\
D_{pp,0}^{\text{halo}} &= \times 10^{-7} (\text{GeV}/c)^2 \, \text{yr}^{-1} \, .
\end{align}

To match the synchrotron emission, we changed $B_0$ to $10 \,
\mu\text{G}$ and $z_0$ to $2 \, \text{kpc}$, keeping $\rho_0 = 5 \,
\text{kpc}$.

In Tbl.~\ref{tbl1}, we have summarised the most important parameters
for the three setups.

%------------------------------------------------------------------------------------------------------------------------------------------
%------------------------------------------------------------------------------------------------------------------------------------------
%------------------------------------------------------------------------------------------------------------------------------------------
\section{Results}
\label{sec:results}

%------------------------------------------------------------------------------------------------------------------------------------------
%------------------------------------------------------------------------------------------------------------------------------------------
\subsection{Model 1: isotropic diffusion}

\begin{figure*}[tbh]
\centering
\includegraphics[scale=1]{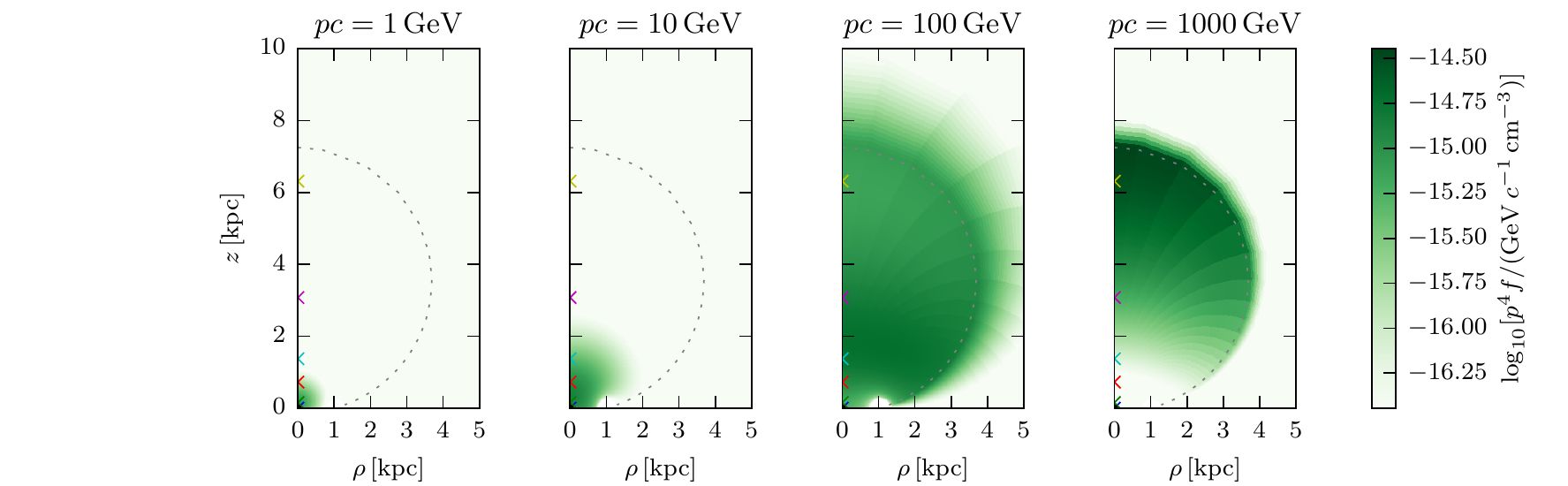}
\includegraphics[scale=1]{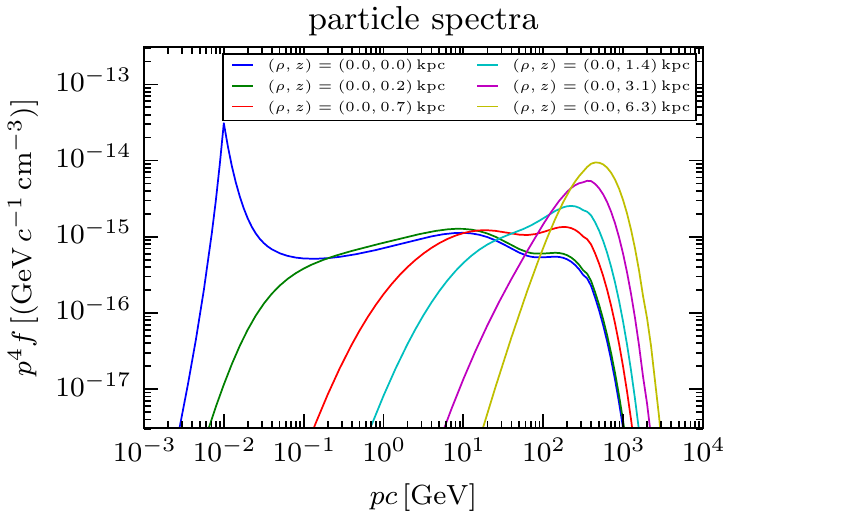}
\caption{\textbf{Top:} The distribution of electron energy $p^4 f \sim
  p^2 \psi \sim E^2 n$ at momenta $pc = {1, 10, 10^2, 10^3} \,
  \text{GeV}$ for model 1 at time $t = 2.4 \times 10^7 \,
  \text{yr}$. The black star marks the position of the source where
  electrons get steadily injected with a momentum $pc = 10^{-2} \,
  \text{GeV}$ and the dashed circle is the shock position at $t = 2.4
  \times 10^7 \, \text{yr}$. The coloured crosses mark the positions
  for which the electron spectra are shown in the bottom panel of this
  figure. \textbf{Bottom:} The spectra $p^4 f $ at time $t = 2.4
  \times 10^7 \, \text{yr}$ for the six positions marked by the
  crosses in the top panel of this figure.}
\label{fig:results_setup1a_electrons}
\end{figure*}

\begin{figure*}[tbh]
\centering
\includegraphics[scale=1]{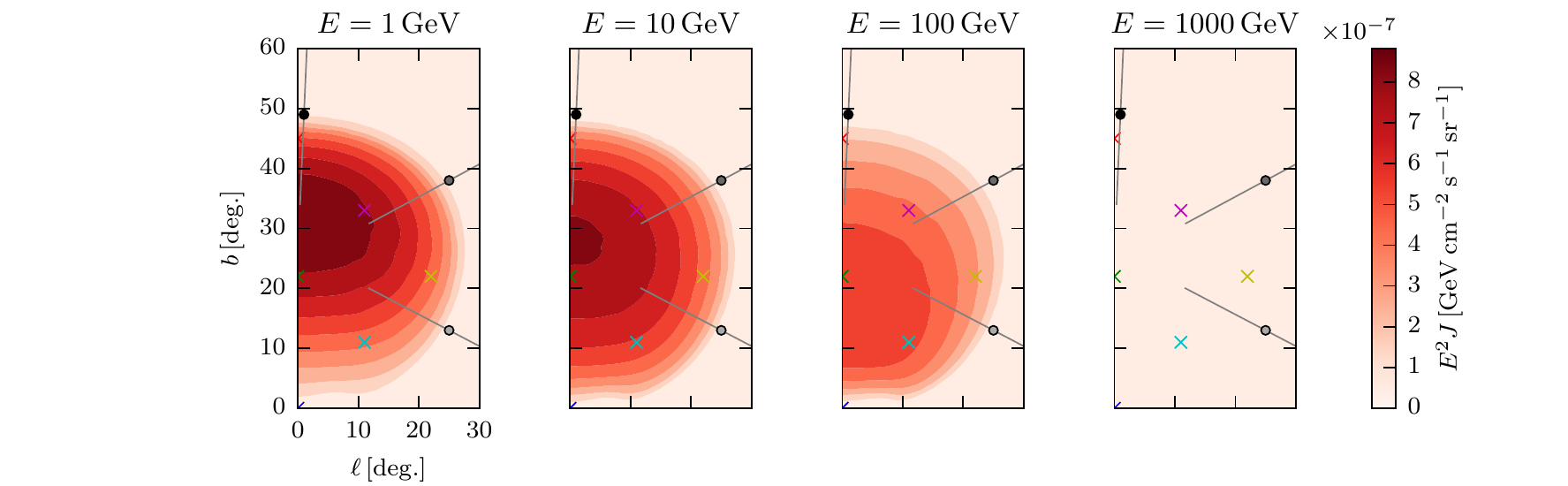}
\includegraphics[scale=1]{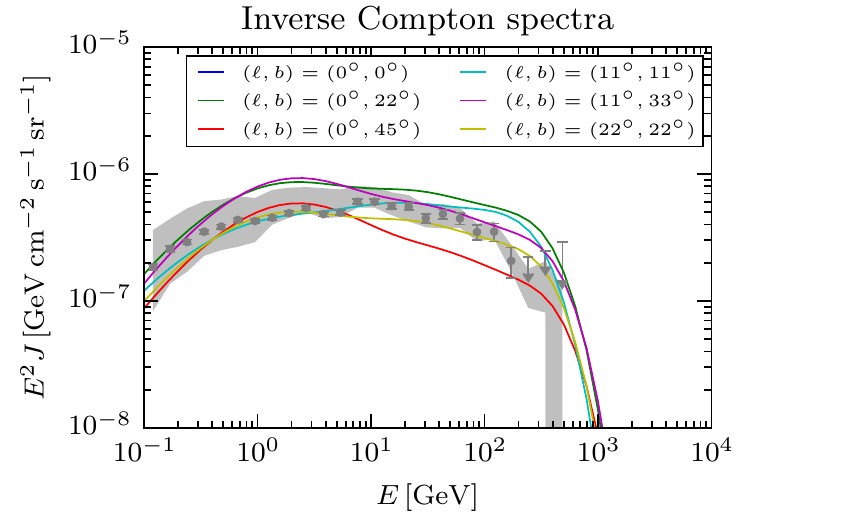}
\includegraphics[scale=1]{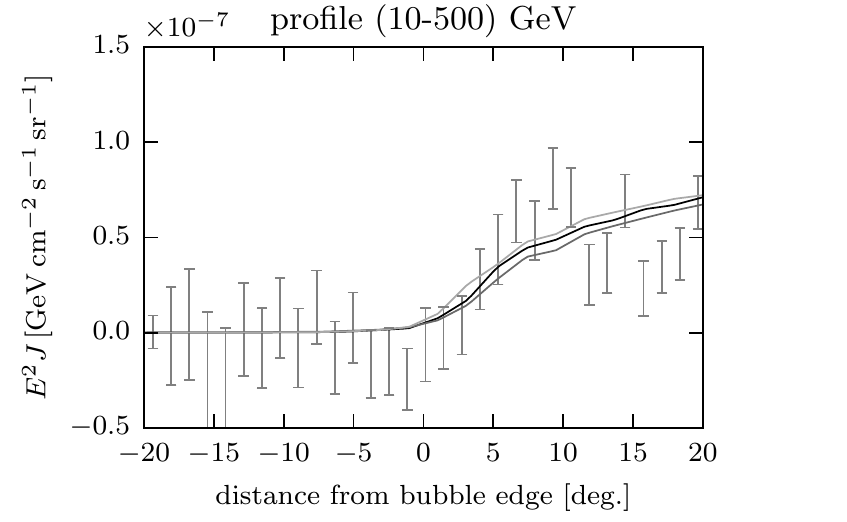}
\caption{\mbox{\textbf{Top:}} Gamma--ray sky maps at $E = 1$, $10$,
  $10^2$ and $10^3 \, \text{GeV}$ for model 1 (homogeneous bubble,
  isotropic diffusion) at time $t = 2.4 \times 10^7 \, \text{yr}$. The
  coloured crosses show the directions for which the gamma--ray
  spectra are shown in the bottom left panel of this figure. The three
  dots mark the directions where the gradient angular directions are
  computed and the lines show these directions. The gamma--ray profile
  along these is shown in the bottom right panel of this figure.
  \mbox{\textbf{Bottom left:}} Gamma--ray spectra in the directions
  marked by the crosses in the top panel of this figure. The data
  points are from~\citet{Fermi-LAT:2014sfa}, showing the statistical
  errors only.  \mbox{\textbf{Bottom right:}} Angular profiles along
  the directions shown in the top panel of this figure for gamma--rays
  in the energy range $10-500 \, \text{GeV}$.  The data points are
  again from~\citet{Fermi-LAT:2014sfa}.}
\label{fig:results_setup1a_gammas}
\end{figure*}

In the top panels of Fig.~\ref{fig:results_setup1a_electrons}, we show
the distribution of electron energy $p^4 f \sim p^2 \psi \sim E^2 n$
as a function of position at energies $pc = 1$, $10$, $10^2$ and $10^3
\, \text{GeV}$ and at time $t = 2.4 \times 10^7 \, \text{yr}$. The
distribution of electrons at GeV energies is very much confined to the
surroundings of the Galactic centre, whereas at $100 \, \text{GeV}$
and $1 \, \text{TeV}$, the distribution extends up to and beyond the
shock. This is due to the fact that high energy electrons have a
larger diffusion coefficient and have thus travelled further from the
source at the Galactic centre while being further accelerated. At $1
\, \text{TeV}$, one can also make out the effect of shock acceleration
which is strongest at the top of the bubble where the advection speed
is highest, cf.\ eq.~\ref{eqn:Vu}. This is leading to a higher
electron energy closer to the shock which will help with producing the
flat intensity profile in gamma--rays.

The bottom panel of Fig.~\ref{fig:results_setup1a_electrons} shows the
electron spectra $p^4 f$ for six positions in the bubbles, marked by
the crosses in the top panels. It can be seen that for $z \gtrsim 1 \,
\text{kpc}$, the electron spectrum is very steep, $f \sim p^{-1}$. The
spectral index lies between those predicted for a steady--state
situation without ($f \sim p^{0}$) and with ($f \sim p^{-4}$)
efficient particle escape (cf., e.g.~\citet{Stawarz:2008sp}). The
electron energy $p^4 f$ is peaked at a few hundred GeV. This energy
scale is set by competition between stochastic acceleration and
radiative energy losses, $t_{\text{sa}}(p_{\text{max}}) =
t_{\text{cool}}(p_{\text{max}})$. This leads to a pile--up of
high--energy electrons just below the maximum energy. At lower
energies, the spectrum is much closer to $f \sim p^{-4}$ whereas here,
both diffusion and advection play the role of efficient particle
escape.

In Fig.~\ref{fig:results_setup1a_gammas} we show gamma--ray sky maps
at $E = 1$, $10$, $10^2$ and $10^3 \, \text{GeV}$ and spectra at six
different positions in the bubbles. On average, the spectrum nicely
reproduces the measurements by
\textit{Fermi}--LAT~\citep{Fermi-LAT:2014sfa}. In the bottom right
panel, we also show the flux profiles along the gradient directions
indicated in the sky maps (upper panels). It can be seen that even in
this simple setup, the flux is increasing within $\sim 10^{\circ}$ of
the bubble edge, but visually it appears in the sky maps that the
bubbles have still edges that are still to soft.

%------------------------------------------------------------------------------------------------------------------------------------------
%------------------------------------------------------------------------------------------------------------------------------------------
\subsection{Model 2: anisotropic diffusion}

\begin{figure*}[!thb]
\centering
\includegraphics[scale=1]{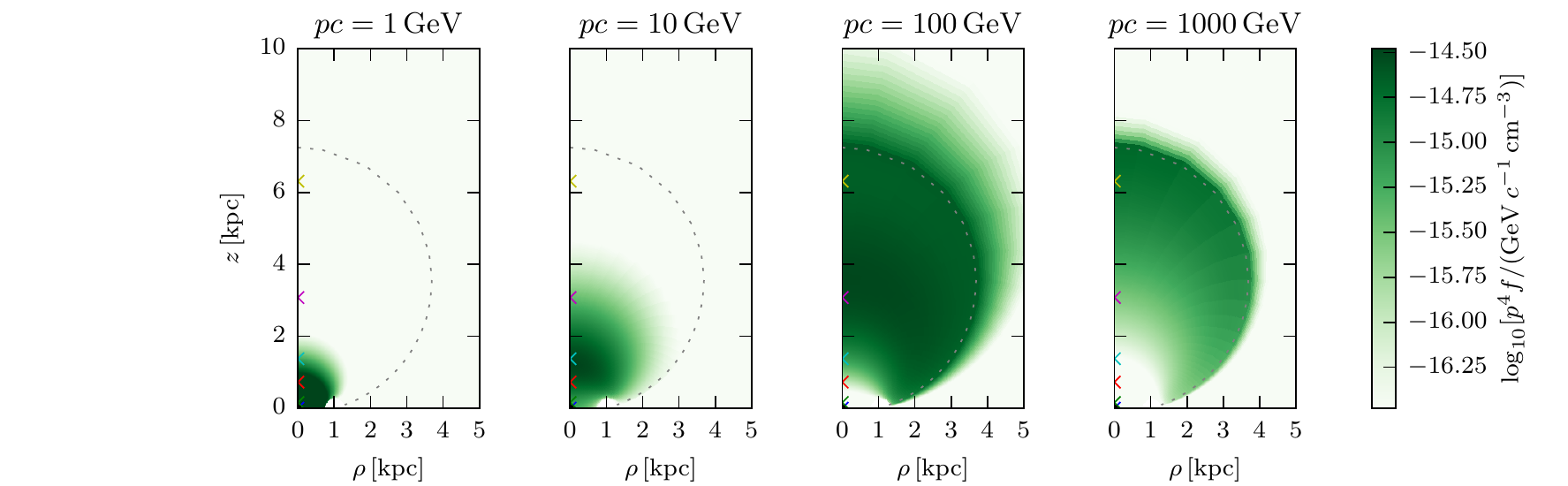}
\includegraphics[scale=1]{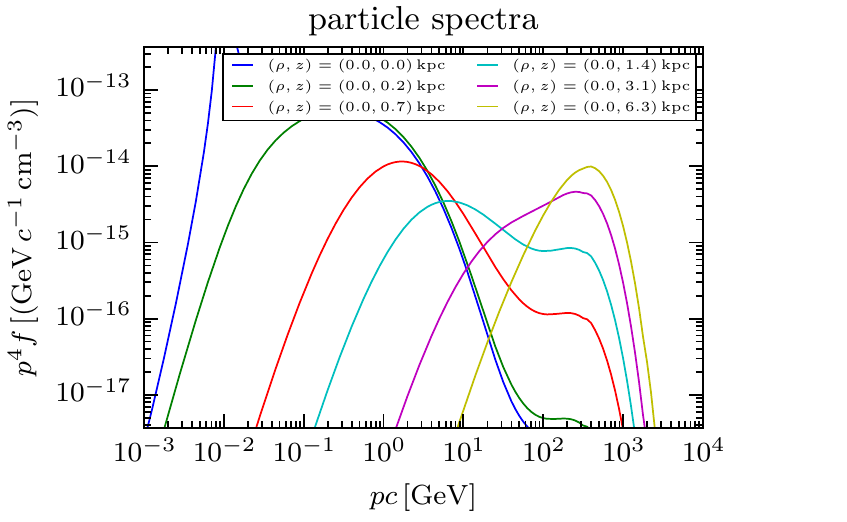}
\caption{Same as Fig.~\ref{fig:results_setup1a_electrons}, but for
  model 2 (homogeneous bubble, anisotropic diffusion) and at time $t =
  2.4 \times 10^7 \, \text{yr}$.}
\label{fig:results_setup2_CRs}
\includegraphics[scale=1]{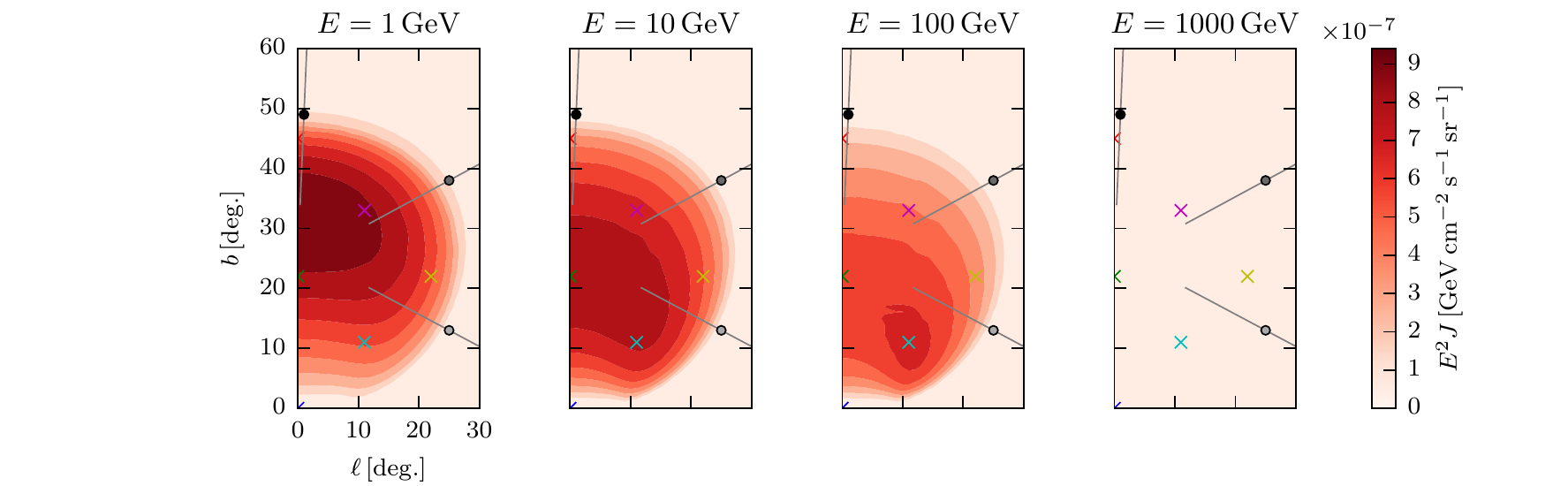}
\includegraphics[scale=1]{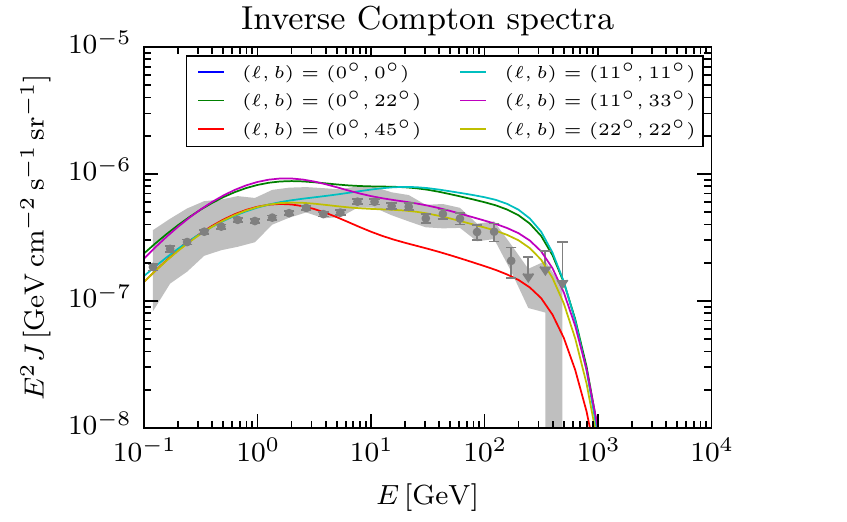}
\includegraphics[scale=1]{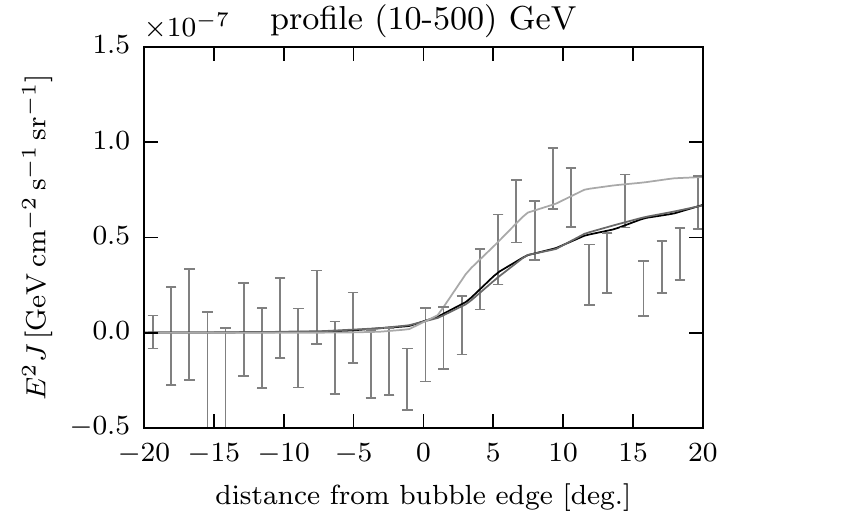}
\caption{Same as Fig.~\ref{fig:results_setup1a_gammas}, but for model
  2 (homogeneous bubble, anisotropic diffusion) and at time $t = 2.4
  \times 10^7 \, \text{yr}$.}
\label{fig:results_setup2_gammas}
\end{figure*}

Looking at Fig.~\ref{fig:results_setup2_CRs}, we see that the
morphology of the electron energy ($p^4 f$) distribution at any one
energy has not changed much with respect to the first model, but that
the low--energy spectrum is much softer. Most of the electron energy
is thus in below--GeV electrons which stay close to the Galactic
centre. This can be understood as parallel diffusion is now faster
than perpendicular diffusion and therefore fewer particles get
transported out into the bubble.

The gamma--ray maps, spectra and profiles,
cf.~\ref{fig:results_setup2_gammas}, look basically the same, as the
high--energy spectrum and morphology are almost unchanged.

%------------------------------------------------------------------------------------------------------------------------------------------
%------------------------------------------------------------------------------------------------------------------------------------------
\subsection{Model 3: anisotropic diffusion and turbulent shell}

\begin{figure*}[!thb]
\centering
\includegraphics[scale=1]{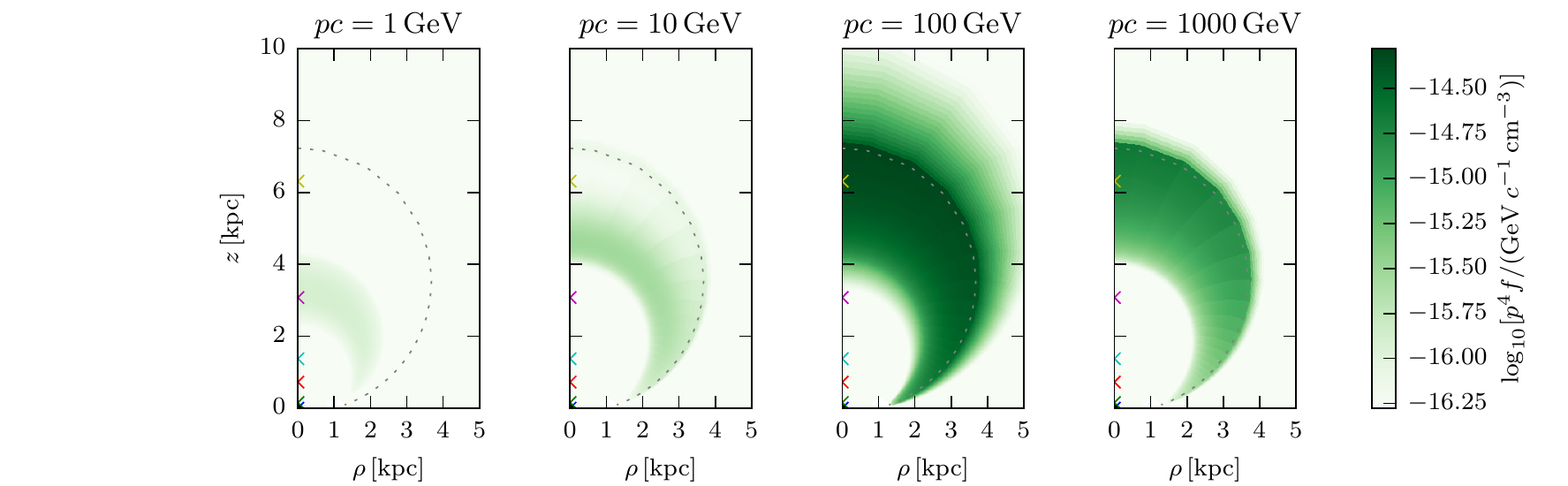}
\includegraphics[scale=1]{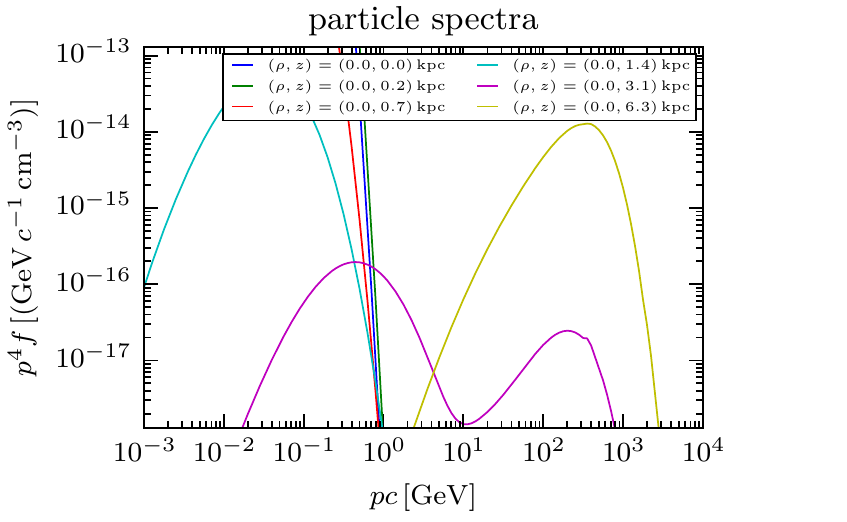}
\caption{Same as Fig.~\ref{fig:results_setup1a_electrons}, but for
  model 3 (bubble with shell, anisotropic diffusion) and at time $t =
  2.4 \times 10^7 \, \text{yr}$.}
\label{fig:results_setup3_CRs}
\includegraphics[scale=1]{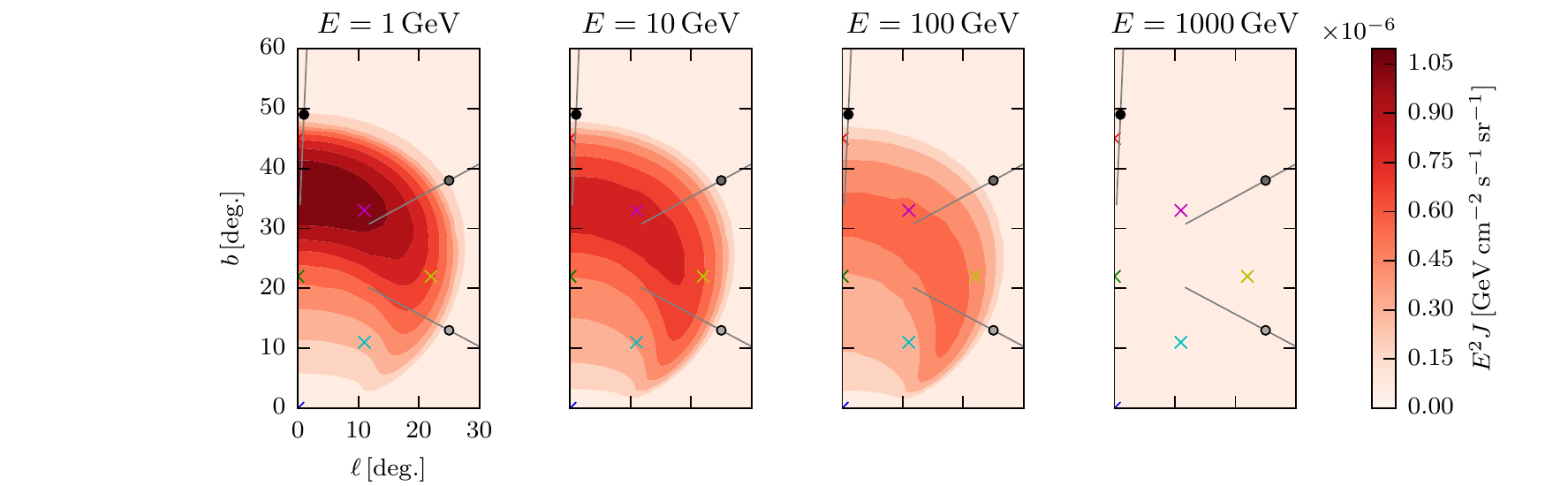}
\includegraphics[scale=1]{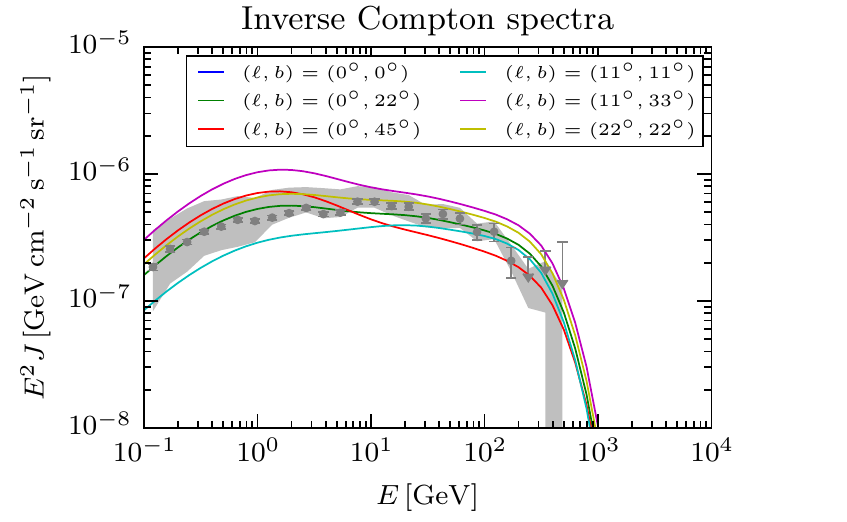}
\includegraphics[scale=1]{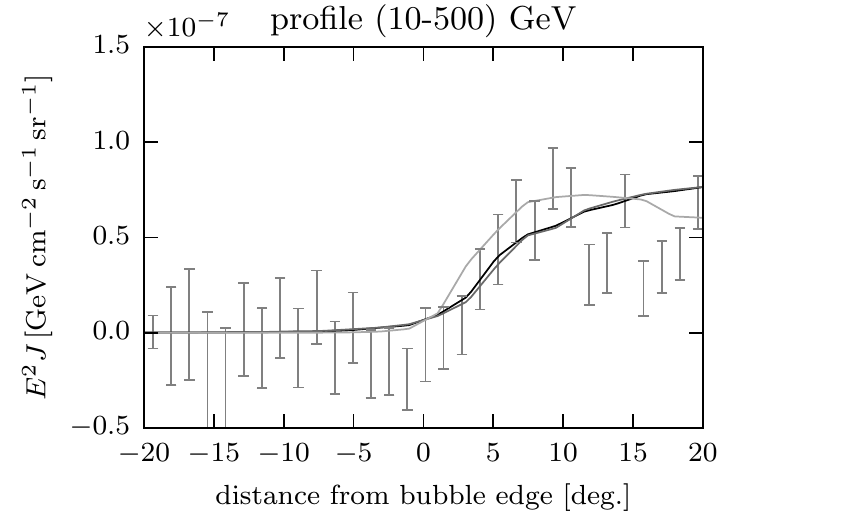}
\caption{Same as Fig.~\ref{fig:results_setup1a_gammas}, but for model
  3 (bubble with shell, anisotropic diffusion) and at time $t = 2.4
  \times 10^7 \, \text{yr}$.}
\label{fig:results_setup3_gammas}
\end{figure*}

Fig.~\ref{fig:results_setup3_CRs} shows a marked difference with
respect to the other setups: The energy range around $1$ and $10 \,
\text{GeV}$ is almost devoid of electrons. At lower energies, the
spectrum is very soft and peaked. At high energies, the picture is
again very similar to models 1 and 2. The morphology at high energies
is also different, in that electrons are only present in the shell
where stochastic acceleration is efficient.

The spectrum is essentially due to the shell geometry assumed in model
3: We recall that due to the scaling of the momentum diffusion
coefficient with turbulence, stochastic acceleration is most efficient
in the shell. In addition, perpendicular diffusion is very much
suppressed inside the bubbles and in the halo. Therefore, only
electrons which were advected into the shell at early times are being
stochastically accelerated. Those that did not reach the shell at
early times will not be able to catch up with the shell through
advection or through diffusion. This constitutes essentially an
selection mechanism that limits the number of electron injected into
stochastic acceleration.

For the production of gamma--rays, the low energy electrons do not
matter.  The edges of the bubbles in gamma--rays are now sharper than
before, a consequence of high--energy electrons only being present in
the shell. Comparing the sky maps at $1$ and $10 \, \text{GeV}$, it is
apparent that the morphology is spectrally rather uniform, as is in
fact observed~\citep{Su:2010qj,Hooper:2013rwa,Fermi-LAT:2014sfa}. This
is even more obvious from the bottom panel of
Fig.~\ref{fig:results_setup3_gammas}.

\begin{figure*}
\includegraphics[width=\textwidth]{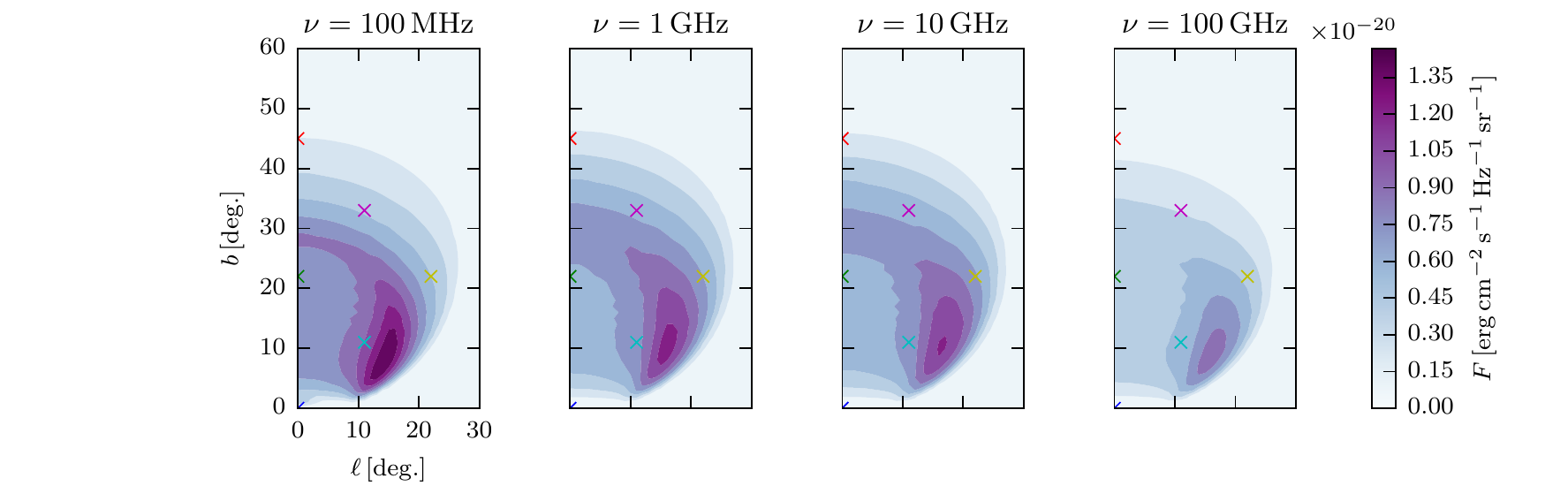}
\centering
\includegraphics[width=0.5\textwidth]{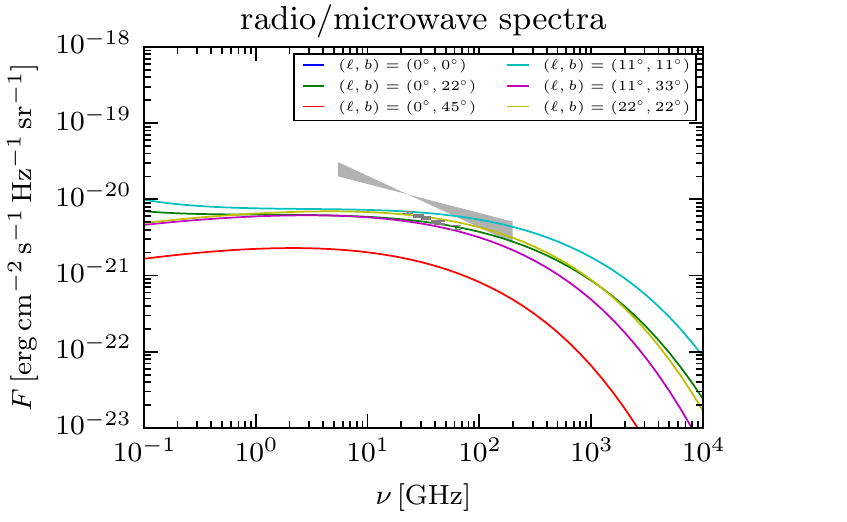}
\caption{\textbf{Top:} Sky maps of the synchrotron intensity at $0.1$,
  $1$, $10$ and $100 \, \text{GHz}$ for model 3 (anisotropic diffusion
  and turbulent shell) at time $t = 2.4 \times 10^7 \, \text{yr}$,
  produced by the same electrons shown in
  Fig.~\protect\ref{fig:results_setup3_gammas}. \textbf{Bottom:}
  Synchrotron spectra for the different directions indicated in the
  legend and marked by the crosses in the lower panel of
  Fig.~\protect\ref{fig:results_setup3_gammas}. The data points are
  from~\citet{Ade:2012nxf} and the bow-tie is
  from~\citet{Dobler:2007wv}.}
\label{fig:model3_haze_synch}
\end{figure*}

Having successfully reproduced the gamma--ray emission, we now ask
whether the electrons could also explain the Galactic microwave
haze~\citep{Dobler:2007wv,Ade:2012nxf}. In
Fig.~\ref{fig:model3_haze_synch}, we show the synchrotron sky maps at
$\nu = \{0.1, 1, 10, 10^2 \} \, \text{GHz}$ and the synchrotron
spectra at various positions in the sky. One can see that the general
morphological characteristics of the microwave haze can be reproduced.
Note that the enhanced emission around the direction $(\ell, b) = (\pm
15, \pm 10)$ would likely be obscured by conventional diffuse
synchrotron emission from the Galaxy.) The synchrotron emission shows
a relatively sharp edge in longitude, but decreases rather smoothly
beyond $\ell = \pm 20^{\circ}$ as observed~\citep{Ade:2012nxf}. The
computed spectra nicely match the spectral index of $-0.5$ observed at
a few tens of GHz, with a hardening below a few GHz; this can explain
why no radio counterpart of the microwave haze has been observed (see,
however~\citep{Carretti:2013sc}). Note that the data points in
Fig.~\ref{fig:model3_haze_synch} are the \emph{average} spectrum of
the haze and should thus be compared to the \emph{average} of the
model lines.

%------------------------------------------------------------------------------------------------------------------------------------------
%------------------------------------------------------------------------------------------------------------------------------------------
%------------------------------------------------------------------------------------------------------------------------------------------
\section{Summary and conclusion}
\label{sec:summary_conclusions}

\begin{enumerate}
\item We have reviewed the observations of the Fermi bubbles in
  gamma-rays relevant for the modeling of the emission, transport and
  acceleration processes with particular focus on their puzzling
  morphological and spectral properties; namely the constancy of their
  surface brightness with abrupt edges and relatively uniform hard
  spectra. We discussed X-ray emission and UV absorption line
  observations which give (somewhat conflicting) bounds on the outflow
  velocities in the bubbles and include in our discussion the
  observations of so-called microwave haze. We reviewed briefly some
  of the models proposed for production of the bubbles including MHD
  simulation, jet and Galactic wind models.
\item Our main focus, however, is the acceleration of particles, their
  transport and emission characteristics with the primary goal of
  explaining the puzzling morphological and spectral
  characteristics. We present arguments in favor of leptonic
  vs.\ hadronic model and develop kinetic equations to described the
  acceleration and transport of electrons throughout the bubbles. We
  include effects of stochastic acceleration by turbulence and those
  of a low Mach number shock with special attention to momentum and
  spatial diffusion coefficient in the magnetized medium of the
  bubble. We also include the effects of energy loss due to inverse
  Compton and synchrotron processes in a inhomogeneous magnetic and
  soft photon (CMB, infrared and optical/UV) fields.
\item We present results from three different models with similar
  characteristics of the (better understood) loss mechanisms but with
  different assumptions about more uncertain acceleration and other
  transport characteristics. The first model has isotropic spatial
  diffusion in the bubble ($\eta=1$) and anisotropic ($K_{uu} \ll
  K_{vv}$; $\eta=10$) in the halo and higher acceleration rate inside
  the bubble. This model results in surface brightness distribution
  not as uniform as observed and can be ruled out. The second model
  has anisotropic diffusion both inside ($\eta=3.16$) and (even
  stronger) outside ($\eta=31.6$) in the halo. This model results in a
  uniform surface brightness but the bubble edge is not as sharp as
  observed. The third model consists of a relatively thick finite size
  shell expanding into the halo with mildly anisotropic diffusion in
  the shell ($\eta=3.16$) but with a much stronger anisotropy inside
  and outside in the halo ($\eta=31.6$). Acceleration rate is ten
  times higher in the shell than outside. This model produces a
  sharper edge and agrees with the observed spectral distribution in
  gamma-rays and microwave ranges.
\item We conclude that the gamma-ray as well as microwave spectral and
  morphological features of the Fermi bubbles can be reproduced by the
  Inverse Compton and synchrotron emission from electrons accelerated
  by turbulence generated in a mildly supersonic outward flowing
  shell. This finding is strengthening the scenario where the bubbles
  are inflated by a wind powered by star formation or star burst
  activity. Another possibility for inflating the bubbles is a jet
  from past AGN activity at the center of the galaxy. Whether an
  \textit{in situ} acceleration of particles in the jet environment
  can lead to explanation of observed characteristics of the bubbles
  as done by our model would require a separate study. If such a
  future study were to conclude that jet models could not produce the
  observed properties, it would strengthen the above conclusions based
  on our current study.\end{enumerate}

%------------------------------------------------------------------------------------------------------------------------------------------
%------------------------------------------------------------------------------------------------------------------------------------------
%------------------------------------------------------------------------------------------------------------------------------------------

\begin{acknowledgements}
The authors are grateful to Anna Franckowiak and Dmitry Malyshev for
continued discussion. This work was supported by Danmarks
Grundforskningsfond under grant no.\ 1041811001. PM was further
supported by DoE contract DE-AC02-76SF00515 and a KIPAC Kavli
Fellowship. This research was funded in part by NASA through
\textit{Fermi} Guest Investigator grant NNH13ZDA001N.
\end{acknowledgements}

\bibliographystyle{aa}
\bibliography{fermi2}

\begin{thebibliography}{49}
\expandafter\ifx\csname natexlab\endcsname\relax\def\natexlab#1{#1}\fi

\bibitem[{Ackermann {et~al.}(2014)}]{Fermi-LAT:2014sfa}
Ackermann, M. {et~al.} 2014, Astrophys. J., 793, 64

\bibitem[{Ackermann {et~al.}(2017)}]{TheFermi-LAT:2017vmf}
Ackermann, M. {et~al.} 2017, Astrophys. J., 840, 43

\bibitem[{Ade {et~al.}(2013)}]{Ade:2012nxf}
Ade, P. A.~R. {et~al.} 2013, Astron. Astrophys., 554, A139

\bibitem[{{Blandford} \& {Eichler}(1987)}]{1987PhR...154....1B}
{Blandford}, R. \& {Eichler}, D. 1987, Phys.\ Rep., 154, 1

\bibitem[{Blumenthal \& Gould(1970)}]{Blumenthal:1970gc}
Blumenthal, G.~R. \& Gould, R.~J. 1970, Rev. Mod. Phys., 42, 237

\bibitem[{Carretti {et~al.}(2013)Carretti, Crocker, Staveley-Smith, Haverkorn,
  Purcell, Gaensler, Bernardi, Kesteven, \& Poppi}]{Carretti:2013sc}
Carretti, E., Crocker, R.~M., Staveley-Smith, L., {et~al.} 2013, Nature, 493,
  66

\bibitem[{Casandjian \& Grenier(2009)}]{Casandjian:2009wq}
Casandjian, J.-M. \& Grenier, I. 2009 [\eprint[arXiv]{0912.3478}]

\bibitem[{Cheng {et~al.}(2014)Cheng, Chernyshov, Dogiel, \& Ko}]{Cheng:2014nva}
Cheng, K.~S., Chernyshov, D.~O., Dogiel, V.~A., \& Ko, C.~M. 2014, Astrophys.
  J., 790, 23

\bibitem[{Cheng {et~al.}(2015{\natexlab{a}})Cheng, Chernyshov, Dogiel, \&
  Ko}]{Cheng:2014lca}
Cheng, K.~S., Chernyshov, D.~O., Dogiel, V.~A., \& Ko, C.~M.
  2015{\natexlab{a}}, Astrophys. J., 799, 112

\bibitem[{Cheng {et~al.}(2015{\natexlab{b}})Cheng, Chernyshov, Dogiel, \&
  Ko}]{Cheng:2015zda}
Cheng, K.~S., Chernyshov, D.~O., Dogiel, V.~A., \& Ko, C.~M.
  2015{\natexlab{b}}, Astrophys. J., 804, 135

\bibitem[{Cheng {et~al.}(2011)Cheng, Chernyshov, Dogiel, Ko, \&
  Ip}]{Cheng:2011xd}
Cheng, K.~S., Chernyshov, D.~O., Dogiel, V.~A., Ko, C.~M., \& Ip, W.~H. 2011,
  Astrophys. J., 731, L17

\bibitem[{Cheng {et~al.}(2012)Cheng, Chernyshov, Dogiel, Ko, Ip, \&
  Wang}]{Cheng:2011tx}
Cheng, K.~S., Chernyshov, D.~O., Dogiel, V.~A., {et~al.} 2012, Astrophys. J.,
  746, 116

\bibitem[{{Crank} {et~al.}(1947){Crank}, {Nicolson}, \&
  {Hartree}}]{1947PCPS...43...50C}
{Crank}, J., {Nicolson}, P., \& {Hartree}, D.~R. 1947, Proceedings of the
  Cambridge Philosophical Society, 43, 50

\bibitem[{Crocker \& Aharonian(2011)}]{Crocker:2010dg}
Crocker, R.~M. \& Aharonian, F. 2011, Phys. Rev. Lett., 106, 101102

\bibitem[{Crocker {et~al.}(2014)Crocker, Bicknell, Carretti, Hill, \&
  Sutherland}]{Crocker:2013mna}
Crocker, R.~M., Bicknell, G.~V., Carretti, E., Hill, A.~S., \& Sutherland,
  R.~S. 2014, Astrophys. J., 791, L20

\bibitem[{Crocker {et~al.}(2015)Crocker, Bicknell, Taylor, \&
  Carretti}]{Crocker:2014fla}
Crocker, R.~M., Bicknell, G.~V., Taylor, A.~M., \& Carretti, E. 2015,
  Astrophys. J., 808, 107

\bibitem[{Dobler \& Finkbeiner(2008)}]{Dobler:2007wv}
Dobler, G. \& Finkbeiner, D.~P. 2008, Astrophys. J., 680, 1222

\bibitem[{Dobler {et~al.}(2010)Dobler, Finkbeiner, Cholis, Slatyer, \&
  Weiner}]{Dobler:2009xz}
Dobler, G., Finkbeiner, D.~P., Cholis, I., Slatyer, T.~R., \& Weiner, N. 2010,
  Astrophys. J., 717, 825

\bibitem[{Finkbeiner(2004)}]{Finkbeiner:2003im}
Finkbeiner, D.~P. 2004, Astrophys. J., 614, 186

\bibitem[{Fujita {et~al.}(2013)Fujita, Ohira, \& Yamazaki}]{Fujita:2013jda}
Fujita, Y., Ohira, Y., \& Yamazaki, R. 2013, Astrophys. J., 775, L20

\bibitem[{Fujita {et~al.}(2014)Fujita, Ohira, \& Yamazaki}]{Fujita:2014oda}
Fujita, Y., Ohira, Y., \& Yamazaki, R. 2014, Astrophys. J., 789, 67

\bibitem[{Guo \& Mathews(2012)}]{Guo:2011eg}
Guo, F. \& Mathews, W.~G. 2012, Astrophys. J., 756, 181

\bibitem[{Hooper \& Slatyer(2013)}]{Hooper:2013rwa}
Hooper, D. \& Slatyer, T.~R. 2013, Phys. Dark Univ., 2, 118

\bibitem[{Kataoka {et~al.}(2015)Kataoka, Tahara, Totani, Sofue, Inoue,
  Nakashima, \& Cheung}]{Kataoka:2015dla}
Kataoka, J., Tahara, M., Totani, T., {et~al.} 2015, Astrophys. J., 807, 77

\bibitem[{Kataoka {et~al.}(2013)}]{Kataoka:2013tma}
Kataoka, J. {et~al.} 2013, Astrophys. J., 779, 57

\bibitem[{Keshet \& Gurwich(2017)}]{Keshet:2016fbq}
Keshet, U. \& Gurwich, I. 2017, Astrophys. J., 840, 7

\bibitem[{Lacki(2014)}]{Lacki:2013zsa}
Lacki, B.~C. 2014, MNRAS, 444, L39

\bibitem[{Langner({2004})}]{Langner:2004zz}
Langner, U.~W. {2004}, PhD thesis, {Potchefstroom University, South Africa}

\bibitem[{Mertsch \& Sarkar(2010)}]{Mertsch:2010ga}
Mertsch, P. \& Sarkar, S. 2010, JCAP, 1010, 019

\bibitem[{Mertsch \& Sarkar(2011)}]{Mertsch:2011es}
Mertsch, P. \& Sarkar, S. 2011, Phys. Rev. Lett., 107, 091101

\bibitem[{Miller \& Bregman(2016)}]{Miller:2016chr}
Miller, M.~J. \& Bregman, J.~N. 2016, Astrophys. J., 829, 9

\bibitem[{Moskalenko \& Strong(1998)}]{Moskalenko:1997gh}
Moskalenko, I.~V. \& Strong, A.~W. 1998, Astrophys. J., 493, 694

\bibitem[{Mou {et~al.}(2014)Mou, Yuan, Bu, Sun, \& Su}]{Mou:2014pea}
Mou, G., Yuan, F., Bu, D., Sun, M., \& Su, M. 2014, Astrophys. J., 790, 109

\bibitem[{Mou {et~al.}(2015)Mou, Yuan, Gan, \& Sun}]{Mou:2015wxa}
Mou, G., Yuan, F., Gan, Z., \& Sun, M. 2015, Astrophys. J., 811, 37

\bibitem[{Orlando \& Strong(2013)}]{Orlando:2013ysa}
Orlando, E. \& Strong, A. 2013, Mon. Not. Roy. Astron. Soc., 436, 2127

\bibitem[{Petrosian(2012)}]{Petrosian:2012ba}
Petrosian, V. 2012, Space Sci. Rev., 173, 535

\bibitem[{Porter \& Strong(2005)}]{Porter:2005qx}
Porter, T.~A. \& Strong, A.~W. 2005, in {Proceedings, 29th International Cosmic
  Ray Conference (ICRC 2005): Pune, India, August 3-11, 2005}, Vol.~4, 77--80

\bibitem[{Sarkar {et~al.}(2015)Sarkar, Nath, \& Sharma}]{Sarkar:2015xta}
Sarkar, K.~C., Nath, B.~B., \& Sharma, P. 2015, Mon. Not. Roy. Astron. Soc.,
  453, 3827

\bibitem[{Sasaki {et~al.}(2015)Sasaki, Asano, \& Terasawa}]{Sasaki:2015nta}
Sasaki, K., Asano, K., \& Terasawa, T. 2015, Astrophys. J., 814, 93

\bibitem[{Stawarz \& Petrosian(2008)}]{Stawarz:2008sp}
Stawarz, L. \& Petrosian, V. 2008, Astrophys. J., 681, 1725

\bibitem[{Su \& Finkbeiner(2012)}]{Su:2012gu}
Su, M. \& Finkbeiner, D.~P. 2012, Astrophys. J., 753, 61

\bibitem[{Su {et~al.}(2010)Su, Slatyer, \& Finkbeiner}]{Su:2010qj}
Su, M., Slatyer, T.~R., \& Finkbeiner, D.~P. 2010, Astrophys. J., 724, 1044

\bibitem[{Thoudam(2013)}]{Thoudam:2013eaa}
Thoudam, S. 2013, Astrophys. J., 778, L20

\bibitem[{Trotta {et~al.}(2011)Trotta, Johannesson, Moskalenko, Porter,
  de~Austri, \& Strong}]{Trotta:2010mx}
Trotta, R., Johannesson, G., Moskalenko, I.~V., {et~al.} 2011, Astrophys. J.,
  729, 106

\bibitem[{Yang \& Ruszkowski(2017)}]{Yang:2017tjr}
Yang, H. Y.~K. \& Ruszkowski, M. 2017, Astrophys. J., 850, 2

\bibitem[{Yang {et~al.}(2012)Yang, Ruszkowski, Ricker, Zweibel, \&
  Lee}]{Yang:2012fy}
Yang, H. Y.~K., Ruszkowski, M., Ricker, P.~M., Zweibel, E., \& Lee, D. 2012,
  Astrophys. J., 761, 185

\bibitem[{Yang {et~al.}(2013)Yang, Ruszkowski, \& Zweibel}]{Yang:2013kca}
Yang, H. Y.~K., Ruszkowski, M., \& Zweibel, E. 2013, Mon. Not. Roy. Astron.
  Soc., 436, 2734

\bibitem[{Zubovas {et~al.}(2011)Zubovas, King, \& Nayakshin}]{Zubovas:2011py}
Zubovas, K., King, A.~R., \& Nayakshin, S. 2011, Mon. Not. Roy. Astron. Soc.,
  415, 21

\bibitem[{Zubovas \& Nayakshin(2012)}]{Zubovas:2012bn}
Zubovas, K. \& Nayakshin, S. 2012, Mon. Not. Roy. Astron. Soc., 424, 666

\end{thebibliography}

\end{document}